\documentclass[journal]{IEEEtran}
\usepackage{cite}
\usepackage{amsmath,amssymb,amsfonts}
\usepackage{algorithmic}
\usepackage[linesnumbered,ruled,vlined]{algorithm2e}
\usepackage{graphicx,color}
\usepackage{makecell}
\usepackage{booktabs}
\usepackage{multicol}
\usepackage{CJKutf8}
\usepackage{multirow} 
\usepackage{cite,amsfonts,amssymb,color, mathtools,setspace,comment,float,array}
\usepackage{textcomp}
\usepackage[hidelinks, breaklinks = true]{hyperref} 
\usepackage{amsmath,amssymb,amsfonts}
\usepackage{algorithmic}
\usepackage{graphicx}
\usepackage{textcomp}
\usepackage{array}
\usepackage{makecell}
\usepackage{xcolor}
\usepackage{subcaption}

\begin{document}
\begin{CJK}{UTF8}{gbsn}

\title{New Mid-Band (FR3, 6-24 GHz) XL-MIMO for 6G: Channel Modeling, Algorithm Evaluation, \\and Field Trials}


\author{Haiyang Miao, Jianhua Zhang, Feifei Gao, Pan Tang, Qi Zhen, Mengxue Li, Ping Zhao, Shihao Zhang,\\ Liang Xia, Guangyi Liu


}


\markboth{Journal of \LaTeX\ Class Files,~Vol.~14, No.~8, August~2021}%
{Shell \MakeLowercase{\textit{et al.}}: A Sample Article Using IEEEtran.cls for IEEE Journals}


\maketitle	

\begin{abstract}

The new mid-band (FR3, 6-24 GHz) spectrum is expected to play an important role in future 6G networks by providing a favorable balance among coverage, capacity, and deployment feasibility. Meanwhile, extremely large-scale multiple-input multiple-output (XL-MIMO) has emerged as a key enabling technology to exploit the propagation and spatial multiplexing potential of these frequency bands. Firstly, this paper provides a systematic review of spectrum allocation and standardization activities for new mid-band spectrum, together with the 6G spectrum planning strategies of countries and regions. Secondly, the wideband massive MIMO channel sounder is also introduced, which is specially developed for channel measurements of new mid-band with over a thousand elements. Thirdly, propagation characteristics and channel modeling approaches of four representative XL-MIMO architectures, including co-located, cell-free, and intelligent XL-MIMO, are comprehensively reviewed and analyzed, with particular emphasis on near-field propagation, spatial non-stationarity, and capacity performance. Then, recent advances in channel estimation, beamforming, and artificial-intelligence-assisted signal processing are summarized. In addition, the performance of new mid-band XL-MIMO systems equipped with 1536 and 768 antenna elements is comparatively evaluated. Finally, real communication environment prototype system field trials conducted in the Upper 6 GHz (U6GHz) band are used to investigate practical system performance under realistic deployment conditions. The results indicate that the target signal-to-noise ratio is a critical factor affecting XL-MIMO performance in the U6GHz band. The sufficiently high signal-to-noise ratio can substantially improve peak downlink capacity and enhance the spatial multiplexing gain of large-scale antenna arrays, whereas uplink performance remains relatively constrained and requires further optimization. This work demonstrates the potential of the new mid-band XL-MIMO in enhancing coverage and system capacity, and points out the key challenges such as near-field signal processing, low-complexity transceiver design, and actual array deployment.

\end{abstract}


\IEEEpeerreviewmaketitle

\section{INTRODUCTION}

With each generation of mobile communication, new frequency bands are allocated to meet the growing demand for connectivity, higher data rates, and enhanced network capacity, ensuring that the system evolves to support modern communication needs. 
In June 2023, the 44th meeting of the International Telecommunication Union - Radiocommunication Sector (ITU-R) Working Party 5D (WP 5D) defined the overall objectives and key trends for sixth-generation (6G) mobile networks. At this meeting, six usage scenarios for International Mobile Telecommunications (IMT) for 2030 (IMT-2030) and beyond were proposed \cite{recommendation2023framework}, emphasizing the growing demand for ultra-high data rates and extensive coverage. However, meeting these demands is likely to require the allocation of additional spectrum resources to communication systems. Extremely large-scale multiple-input-multiple-output (XL-MIMO) can be regarded as the further extension of 5G massive MIMO to a higher spatial dimension, i.e., using hundreds or even thousands of antennas at the base station (BS) \cite{huo2023technology}. As a result of the deeper use of space resources, XL-MIMO can significantly increase system capacity and spectrum efficiency \cite{tangxlmimo, zhi2024performance}. 

To achieve the ambitious vision of 6G and support the design of key enabling technologies like XL-MIMO, the fundamental understanding of channel characteristics is indispensable. In particular, exploring the novel XL-MIMO channel characteristics within the mid-band spectrum and developing practicable, accurate channel models are crucial for innovative research and performance evaluation of future communication systems \cite{zhang2023channel, wang_mimo}. The new mid-band XL-MIMO channel exhibits unique new characteristics, such as near-field non-stationarity, which pose new requirements for algorithm evolution. The development of the prototype device provides comprehensive support to address this challenge. On the one hand, it creates conditions for researchers and practitioners to explore the implementation schemes of 6G technology in the new mid-band. On the other hand, it effectively promotes the process of open development.

In recent years, the academic community has conducted a series of studies on XL-MIMO channel characteristics, algorithm, and field trials of prototypes in the new mid-band. 

\subsubsection{Recent work on XL-MIMO channel characteristics}
The wireless channel serves as the transmission medium between the transmitter (Tx) and receiver (Rx). It plays a crucial role in determining the ultimate performance limit of mobile communication systems \cite{molisch2012wireless}. The channel model is a mathematical description of the effects of a communication channel through which wireless signals are propagated. 

Several channel measurement activities were conducted to investigate other characteristics of the XL-MIMO channel. Based on the XL-MIMO channel measurement campaigns, spatial non-stationarity (SnS) has been observed. In \cite{chen2016measurement}, it was discovered that channel gains exhibit random variations in the array, with no clear deterministic trend. Moreover, it was indicated that the K-factor at different positions on the array fluctuates within a certain range. In \cite{yang2022channel}, the spatial non-stationary channel capacity calculation method was proposed. To further identify the SnS characteristics, the work in \cite{Jing2026measurement} established a SnS statistical channel model for XL-MIMO systems based on extensive channel measurements. Besides, a model was proposed based on the dominant multipath propagation mechanism, which is implemented by introducing a novel matrix\cite{yuan2022spatial}. The analyses in \cite{miao2025far} revealed that the spatial non-stationary phenomenon can be more obvious along the array in the near-field region, compared to the far-field region. Besides, the extension channel model is proposed based on the channel model of 3GPP TR 38.901 \cite{3gpp38.901}. The array domain is introduced to characterize the SnS characteristics. 

The results in \cite{yuan2022spatial} and \cite{zheng2022ultra} confirmed the phenomenon of spherical wavefronts.  Based on channel measurements, THz XL-MIMO channels were further characterized for both the far-field and near-field regions, validating near-field propagation features \cite{Wang2024far-}. Furthermore, the unified cross-field path loss model is applicable to both regions. Moreover, the results in \cite{willhammar2020channel} indicate that interactions between the array, the environment, and user movement will affect the degree of channel hardening. In the context of emerging technologies, the authors in \cite{zhou20256g} investigated the reconstruction of environmental objects in the mid-band, validating the potential of Integrated Sensing and Communications (ISAC).

\subsubsection{Recent work on XL-MIMO algorithm}
Channel estimation and beamforming, as the key enabling technology algorithms of XL-MIMO systems, have attracted extensive and in-depth research attention in both academia and industry in recent years. These algorithms are not only the core means to enhance the spectral efficiency and energy efficiency of the system, but also the foundation for achieving high-frequency communication, extremely high spatial resolution and user-level precise beam management.
\begin{itemize}
    \item Channel Estimation
\end{itemize}

Targeting near-field channel estimation in XL-MIMO systems equipped with uniform planar arrays, \cite{Peng2024channelEstimation} introduced two novel algorithms: dual simultaneous orthogonal matching pursuit (DS-OMP) and single iterative OMP (SI-OMP), the latter of which has lower computational complexity. The simulation results showed that the proposed DS-OMP and SI-OMP algorithms can achieve better normalized mean square error (NMSE) performances than the existing methods.

Exploring generative AI applications, \cite{Ye2025GAN} proposed a generative adversarial network (GAN)-based framework explicitly designed to simultaneously estimate line-of-sight (LOS) and non-LOS (NLOS) path components in mixed near-field XL-MIMO channels. Numerical results showed that the proposed method based channel estimation method outperformed the two stage algorithm and the far-field codebook based OMP scheme in the adopted mixed LOS/NLOS XL-MIMO near-field channel model. Furthermore, the proposed method surpassed the Cramer-Rao lower bound (CRLB) when the distance between the transmitter and the receiver was small and the pilot overhead was low.

By transforming this problem into the multidimensional compressed sensing (CS) problem, \cite{Ruan2024low} investigated the near-field channel estimation problem in XL-MIMO systems, and proposed a multidimensional orthogonal matching pursuit relying on parameter refinement (MOMP-PR) algorithm to solve it, achieving near-field channel reconstruction at a lower complexity. The simulation results showed that this algorithm can achieve better NMSE performance than existing CS-based NMSE algorithms.

\begin{itemize}
    \item Beamforming
\end{itemize}

To mitigate the computational burden of XL-MIMO transmission, \cite{Ribeiro2021low} introduced two novel precoding schemes: mean-angle based zero-forcing (MZF) and tensor zero-forcing (TZF). By leveraging plane-wave approximations and strategic user grouping, these methods achieve a low-complexity approximation of the standard ZF precoder. The simulation results showed that both MZF and TZF closely approach the performance of benchmark solutions, providing a highly effective trade-off between computational complexity and system performance.

Providing a comprehensive overview of near-field resource allocation in XL-MIMO systems, \cite{Xu2024resource} evaluated various optimization frameworks, ranging from traditional numerical techniques and machine learning models to emerging AI-generated approaches. By delineating their respective advantages and fundamental limitations, the study highlighted that AI-generated optimization methods are particularly adept at navigating highly complex and dynamically shifting network environments.

The authors in \cite{Xu2023jac} investigated four iterative matrix inversion methods for XL-MIMO systems, with the goal of fast matrix inversion for regularized zero-forcing (RZF) precoding. These four iterative methods were: Gauss-Seidel (GS), Jacobi Over Relaxation (JOR), Conjugate Gradient (CG), and Jacobi-Preconditioning Conjugate Gradient (Jac-PCG) methods. The authors compared these methods from the perspective of spectral efficiency (SE) performance, computational complexity, and convergence speed. The results indicated that compared to the JOR method, the GS method exhibits a higher SE performance and a faster convergence rate. However, the computations of the GS method cannot be parallelized and have high complexity. In contrast, the Jac-PCG method achieves a trade-off between complexity and performance for XL-MIMO systems.

\begin{figure*}[!ht]
    \centering
    \includegraphics[width=0.9\linewidth]{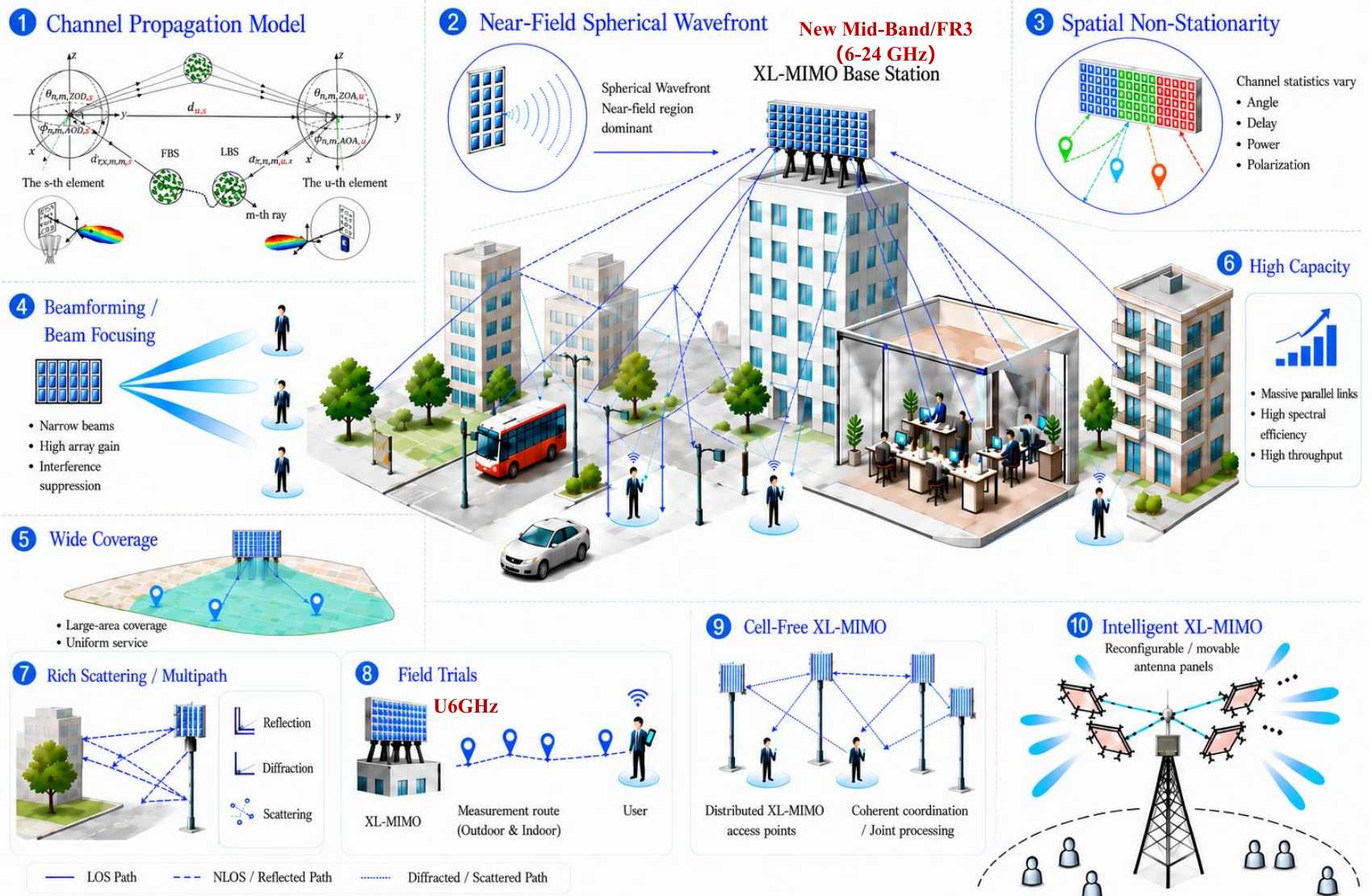}
    \caption{XL-MIMO characteristics and deployment in the new mid-band (FR3) band.}
    \label{fig:Structure}
\end{figure*}

\subsubsection{Recent work on the field trials of prototypes}
Besides, The research, education and industry communities in various countries have actively pursued the development and experimental validation of prototypes in the new mid-band spectrum. 
\begin{enumerate}
    \item \textbf{China}: In June 2025, China Mobile was the first to complete an innovative test and validation of U6GHz (Upper 6 GHz, 6425-7125 MHz) in Zhejiang. Leveraging an industry-leading U6GHz 256 TRx base-station solution, the trial achieved a stable user experience with a peak down-link throughput exceeding 10 Gbps under four-component-carrier aggregation. In August 2025, ZTE Corporation proposed the large-scale array prototype featuring more than a thousand radiating elements for the 6425-7125 MHz frequency band, and further introduced a distributed ultra-large-array architecture for 5G-Advanced low-altitude unmanned aerial vehicle (UAV) communications.
    \item \textbf{America}: In October 2023, Tarana announced that its next-generation fixed wireless technology can operate in the 5925-6425 MHz frequency band. The company positions this technology as a cost-effective and rapidly deployable alternative that outperforms both fiber-to-the-home solutions and satellite-based systems. Later in November 2023, Nextlink proposed a plan to deploy fixed wireless equipment from Cambium Networks in the 6 GHz frequency band. In February 2025, Qualcomm announced its efforts to advance 6G standardization for new mid-band spectrum and to collaborate with Nokia Bell Labs on the development of AI-enhanced network architectures. At Mobile World Congress (MWC) 2026, Qualcomm introduced an end-to-end 6G prototype system operating in the 7 GHz band. The system features 2048 antenna elements and supports 256 transmit–receive channels.
    \item \textbf{Finland}: In September 2025, Nokia conducted a proof-of-concept trial in Oulu using frequencies close to 7 GHz with four test mobile devices. The trial evaluated the performance of these terminals in an urban environment and examined the impact of varying levels of attenuation on signal quality. In February 2026, Nokia submitted an application for a short-term license to conduct prototype equipment testing in the 6925–7125 MHz frequency band, with the objective of advancing 6G technologies and exploring novel wireless communication solutions.
    \item \textbf{Japan}: In July 2025, Rakuten Mobile emphasized at the 3GPP 6G Workshop the importance of fully exploiting the potential of mid-band spectrum and called for the development of new spectrum aggregation techniques to improve mid-band efficiency. Meanwhile, SoftBank, in partnership with Nokia, has conducted Japan’s first outdoor 7 GHz field trial for potential 6G use by deploying three pre-commercial base stations in Tokyo and co-locating them with 3.9 GHz sites for comparative evaluation. In November of the same year, tests conducted by SoftBank demonstrated that the 7 GHz band can provide effective wide-area coverage and maintain robust link stability in dense urban environments.
    \item \textbf{Malaysia}: In September 2023, the operator Maxis, in collaboration with Universiti Malaya, conducted field trials in the 6 GHz frequency band to enhance the capability of mobile networks to support emerging services.
    \item \textbf{Korea}: In March 2026, Samsung completed an outdoor field trial of large-scale MIMO in the 7 GHz band in Seoul, systematically demonstrating the technical feasibility of applying this frequency band to 6G mobile communication systems.
\end{enumerate}

This paper presents a comprehensive study on radio propagation in outdoor environments, conducted at mid-band frequencies using a high-precision channel sounder.  Direct comparisons with 3GPP models are given in this paper, and offer valuable insights into the accuracy of models and how they might need to be revised for greater accuracy and broader applicability. Fig. \ref{fig:Structure} illustrates the characteristics and deployment of XL-MIMO. The key contributions in this paper are as listed:

\begin{itemize}
    \item Section \ref{sec:II} mainly introduces the 6G spectrum requirements and progress. Firstly, we propose a guiding principle of focusing on adjacent frequency bands for spectrum selection. Then, we systematically review the spectrum allocation and standardization progress of the frequency bands sub-6 GHz as well as new mid-band (6-24 GHz), which support the enhanced data transmission capability of future mobile communication systems. In addition, we also highlight and summarize the plans and strategies for the division of 6G spectrum in regions.
\end{itemize} 

\begin{itemize}
    \item Section \ref{sec:III} introduces the large-scale MIMO wideband channel sounder, which is specially designed for devices operating in the new mid-band (6-24 GHz band). In addition, a detailed measurement methodology is described to capture the radio propagation behavior in the outdoor environment and to conduct measurements. 
\end{itemize} 

\begin{itemize}
    \item Section \ref{sec:IV} divides XL-MIMO into four types: co-located, cell-free, sparse, and movable, based on the differences in system architecture and deployment methods. It elaborates on the channel characteristics of each of these four types of XL-MIMO, including spatial characteristics, near-field characteristics, capacity performance, etc., and discusses the corresponding channel models.
\end{itemize}

\begin{itemize}
    \item Section \ref{sec:V} provides a comprehensive review of channel estimation, beamforming scheme design, and processing enabled by artificial intelligence. Due to the extremely large array aperture of the XL-MIMO system, the signal processing scheme will involve extremely high computational complexity. Therefore, algorithms are classified into traditional algorithms, algorithms based on artificial intelligence, and low-complexity signal processing schemes. Most importantly, the performance such as the signal-to-noise ratio of the new mid-band XL-MIMO is compared with 1536 and 768 array elements.
\end{itemize}

\begin{itemize}
    \item Section \ref{sec:VI} discusses the actual performance of XL-MIMO based on field trials in the U6GHz (6425-7125 MHz) band. Firstly, the system architecture, parameter configuration, some performance indicators, and functional implementation of the 1024-element XL-MIMO prototype system are presented. Then, several key performance test results are analyzed and discussed in depth.
\end{itemize}

\section{The Requirement and Progress of 6G New Mid-Band Spectrum}
\label{sec:II}
\begin{figure}[!htbp]
    \centering
    \includegraphics[width=0.45\textwidth]{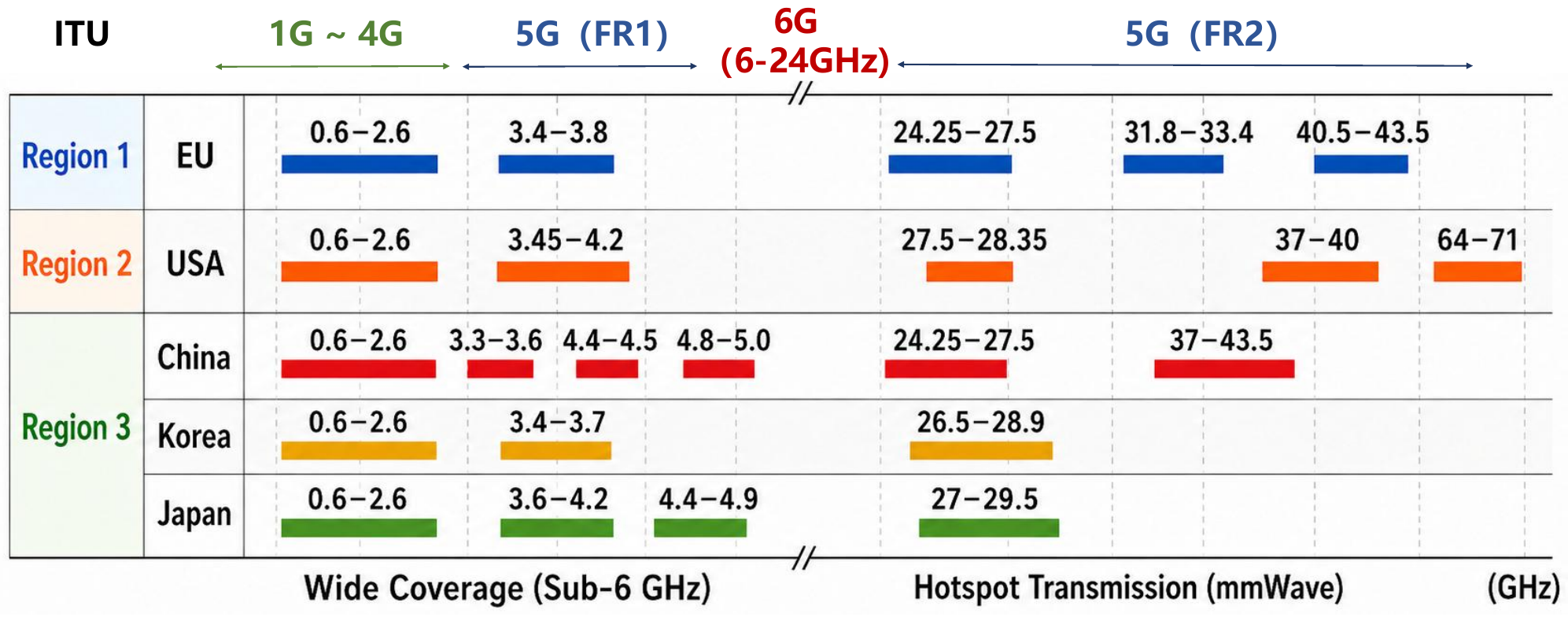}
    \caption{The operating and potential spectrum in ITU regions.}
    \label{fig:figure1}
\end{figure}

The global frequency spectrum is managed by dividing the world into three ITU regions. Fig.~\ref{fig:figure1} illustrates the spectrum allocation for selected countries within these regions \cite{zhang2024new}. From the first generation (1G) to 5G, it is evident that mobile communication frequency bands are allocated within the sub-6 GHz range as well as above 24 GHz. In the upcoming 6G era, collaborative communications across multiple frequency bands are expected to enable efficient utilization of both low- and high-frequency resources, thereby offering significant application potential. However, as the frequency span increases, the variations in channel characteristics become more pronounced, presenting substantial challenges for multi-band wireless networking. Consequently, it is advisable to limit the frequency span and focus on adjacent frequency bands. These challenges highlight the critical importance of carefully selecting frequency bands for 6G development. 

The sub-6 GHz frequency bands in fifth-generation (5G) networks can be reused through spectrum refarming; however, there is also a need to identify wider bandwidth spectrum resources. Currently, 6G development mainly focuses on the new mid-band spectrum (6-24 GHz)\cite{zhang2024new}. In June 2022, the 3rd Generation Partnership Project (3GPP) Radio Access Network (RAN) Plenary approved  a standard modification proposal for the 5925-7125 MHz band, officially including it into 5G-Advanced Release 18. The 2023 World Radiocommunication Conference (WRC-23) designated the 6 GHz band for mobile use in all International Telecommunication Union (ITU) regions and set the agenda for WRC-27. The new WRC cycle will prioritize the following frequency bands for International Mobile Telecommunications (IMT): 4400-4800 MHz, 7125-8400 MHz, and 14.8-15.35 GHz. In December 2023, the 3GPP Technical Specification Group (TSG) RAN Release 19 initiated discussions on research activities within the FR3 band. At the 111th 3GPP TSG-RAN meeting held in Fukuoka, Japan, in March 2026, a research roadmap was established for the maximum channel bandwidth in the 7 GHz band to support enhanced data transmission capabilities for future mobile communication systems.

Countries around the world have initiated strategic plans for the allocation of 6G spectrum. 
\begin{enumerate}
    \item \textbf{Asia}: In July 2023, the Ministry of Industry and Information Technology (MIIT) of China released a new version of the ``Radio Frequency Allocation Regulations of the People's Republic of China," becoming the first in the world to allocate the upper 6 GHz band, from 6425 to 7125 MHz, exclusively for 5G/6G systems. In December 2024, Hong Kong completed the world’s first 6 GHz spectrum auction. China Mobile Hong Kong Limited won the 6570-6670 MHz band, securing a 100 MHz spectrum, while Hong Kong Telecommunications and SmarTone Telecommunications acquired the 6670–6770 MHz and 6925-7025 MHz bands, respectively, for 5G and future 6G mobile communication services. In May 2025, the Ministry of Telecommunications of India allocated the lower 6 GHz band (5925-6425 MHz) as unlicensed spectrum for use by wireless access technologies such as WiFi, while reserving the upper band (6425-7125 MHz) for telecommunication operators in India for 6G spectrum allocation. In May 2026, the MIIT of China approved the use of experimental frequencies for the IMT-2030 (6G) Promotion Group. This approval is intended to support research, development, and test verification of 6G technologies in the 6425-7125 MHz band, with a focus on typical 6G usage scenarios and key performance indicators defined by the ITU.
    \item \textbf{Americas}: In July 2025, the U.S. Federal Communications Commission (FCC) reallocated the 6.425-7.125 GHz frequency band, originally designated for Wi-Fi use, to mobile network operators such as AT\&T, Verizon, and T-Mobile. The US President takes action to win the 6G race and signed a presidential memorandum on December 19, 2025. The document prioritized clearing federal systems from the 7.125-7.4 GHz band for high-power commercial 6G use and began the reallocation process. It also urged the FCC and other organizations to form international industry alliances ahead of the 2027 World Radiocommunication Conference (WRC) to discuss the 7.125-8.4 GHz band for global mobile communications. In February 2026, T-Mobile obtained an experimental authorization from the U.S. FCC to conduct 6G prototype testing at and around its headquarters in Bellevue, Washington. The trial utilizes prototype equipment provided by Ericsson and MediaTek and operates in the 6425–6825 MHz band. The objective of the experiment is to investigate the spectrum performance characteristics of higher mid-band frequencies for potential 6G deployments.
    \item \textbf{Europe}: In November 2024, the EU Radio Spectrum Policy Group (RSPG) released the draft 6G Strategic Vision, noting that higher-frequency spectrum below 7 GHz (including the upper 6 GHz band) and new bands in the 7–15 GHz range will be needed to complement low-band deployments and provide sufficient capacity, particularly in urban and suburban areas. In February 2025, the UK communications regulator Ofcom issued a consultation document on the upper 6 GHz band, proposing that the spectrum above 6 GHz (6425-7125 MHz) be designated for both Wi-Fi and mobile communication services. In a consultation document published in early 2026, the UK Office of Communications (Ofcom) introduced a novel priority-sharing framework for the 6 GHz band. Specifically, high-power Wi-Fi is proposed for deployment in the lower sub-band (5925–6425 MHz), while the upper sub-band (6425–7125 MHz) is partitioned into 160 MHz designated for Wi-Fi priority access and 540 MHz allocated for mobile-network-priority access, thus improving overall spectrum utilization efficiency.
    \item \textbf{Africa}: In 2024, South Africa witnessed intense competition between Wi-Fi and IMT (cellular) operators over spectrum allocation in the 6 GHz band. 
    \item \textbf{Oceania}: In June 2021, New Zealand’s Radio Spectrum Management (RSM) released a consultation on the 6 GHz band plan, addressing potential future usage scenarios for the 5925-7125 MHz frequency range. In June 2024, the Australian Communications and Media Authority (ACMA) released the “Future Use of the Upper 6 GHz Band” Options Paper and initiated a public consultation on planning approaches for the U6GHz.
\end{enumerate}

\section{Mid-Band XL-MIMO Experimental Platform}
\label{sec:III}

\par
\begin{figure}[!htbp]
	\xdef\xfigwd{\columnwidth}
	\setlength{\abovecaptionskip}{0.1 cm}
	\centering
	\begin{tabular}{c}
		\includegraphics[width=0.42\textwidth]{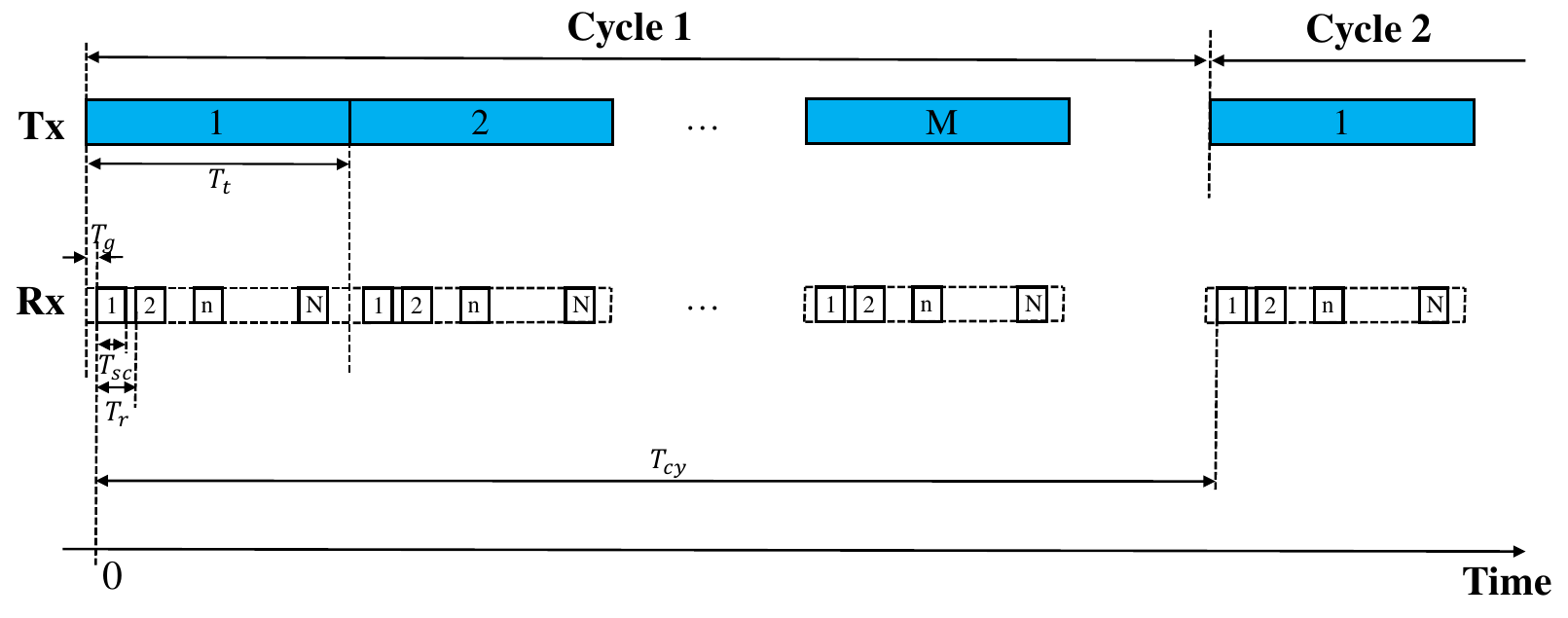}\\
		{\footnotesize\sf (a)} \\[3mm]
		\includegraphics[width=0.42\textwidth]{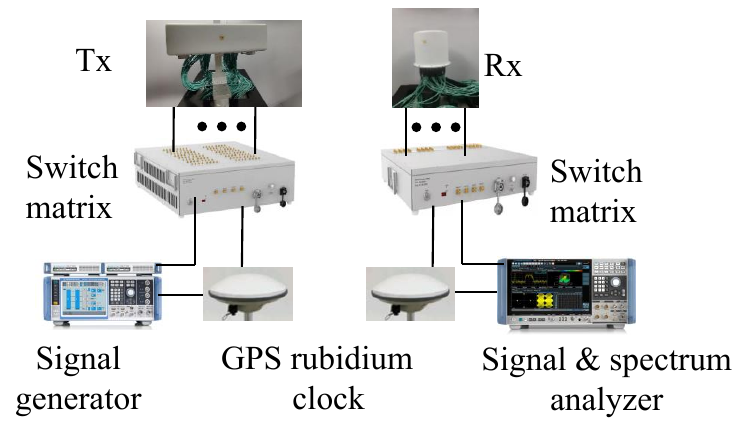}\\
		{\footnotesize\sf (b)} \\[3mm]
        \includegraphics[width=0.42\textwidth]{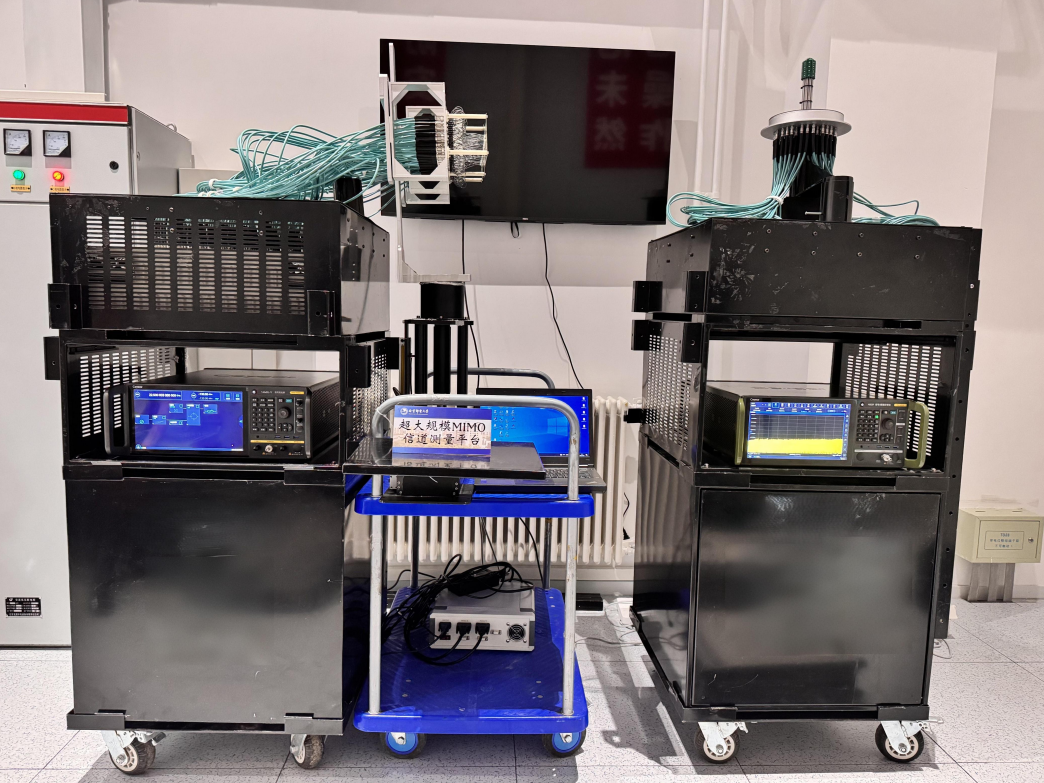}\\
		{\footnotesize\sf (c)} \\
	\end{tabular}
	\caption{ Massive MIMO channel sounder. (a) TDM-MIMO working principle. (b) Schematic diagram of massive MIMO channel measurement platform. (c) Experiment equipment.}
	\label{Figure_mimosounding}
\end{figure}

 Research on channel characteristics and channel modeling is fundamental to every generation of mobile communication systems. Channel measurement is a primary means of understanding real-world wireless propagation environments and acquiring fundamental channel data. As illustrated in Fig. \ref{Figure_mimosounding} and Table \ref{mimosounding}, the mid-band massive MIMO platform employs high-speed electronic switches to complete channel measurements, thus facilitating comprehensive TDM-MIMO channel characterization at medium and low frequencies \cite{Miao2024Measurement}. By integrating pseudo-random (PN) sequence signals with time-division multiplexing antenna switching technology, the system effectively decouples the response of the antenna measurement system itself. $T_t$: transmitting switch duration, the duration of the signal on an array at the transmitting end. $T_r$: duration of the receiving switch. $T_{cy}$: the time of a probe cycle, which must meet $T_{cy}\geq MT_t$. $T_g$: protection interval, which will be reset to zero in the next period. $T_{sc}$: duration of continuous detection of each element at the receiving end. $T_s$: pulse period. In one cycle, each receiving antenna array is detected once. A specified probe duration, $T_t$ contains exactly one detection period, that is, $T_t=NT_{sc}$. As switching takes time in the actual operation process, the protection interval must be set, so the interval between two consecutive detection signals is specified as $T_r$, where $T_r\geq T_{sc}$. 

\par The transmitter end comprises a high-frequency signal generator, control computer, switch matrix, GPS rubidium clock module, feeder, and antenna array. Meanwhile, the receiver end is composed of a MIMO antenna array, feeder, spectrum analyzer, control computer, switch matrix, and GPS rubidium clock module. Fig. \ref{Figure_mimosounding} presents a simplified schematic of the measurement system, including the signal transceiver and the transceiver antenna matrix module.

The transceiver antenna matrix module leverages the GPS reference clock to achieve high-precision synchronization initialization and synchronization unit switching. In addition, the signal transceiver relies on the 10 MHz GPS reference clock to maintain the consistency of the base-band signal of the transmitter, the carrier and I/Q demodulation frequencies at the receiver end, and the timing sequence for time-domain sampling. As a result, signal transmission, antenna switching, and signal acquisition all benefit from precise synchronization. The response of the array antenna is measured, demonstrating an angular resolution less than 2°.

\begin{table}[!htbp]
	\centering
	\caption{Performance of massive MIMO channel sounding platform covering new mid-band}
	\setlength{\tabcolsep}{0.1 mm}
	\label{mimosounding}
	\renewcommand{\arraystretch}{1.5}
	 \setlength{\tabcolsep}{1 mm}
	\begin{tabular}{c|c}
	 \hline \hline
	    Parameter    & Performance  \\ \hline
		Work frequency [GHz]    & 3-16  \\ \hline
		Max bandwidth [GHz]     & 2  \\  \hline
		Number of Tx switch matrix channel &   128  \\   \hline
		Tx switch gain [dB] &  $\geq$ 27    \\   \hline
		Number of Rx switch matrix channel   & 64  \\   \hline
		Rx switch gain [dB] &  $\geq$ 33      \\   \hline
		Antenna angular resolution [$^\circ$]  &    $\leq$ 2  \\   \hline
		Synchronous mode    & GPS rubidium clock\\   \hline \hline
	\end{tabular}
\end{table}

\begin{figure}[htbp]
\centering
\includegraphics[width=0.9\linewidth]{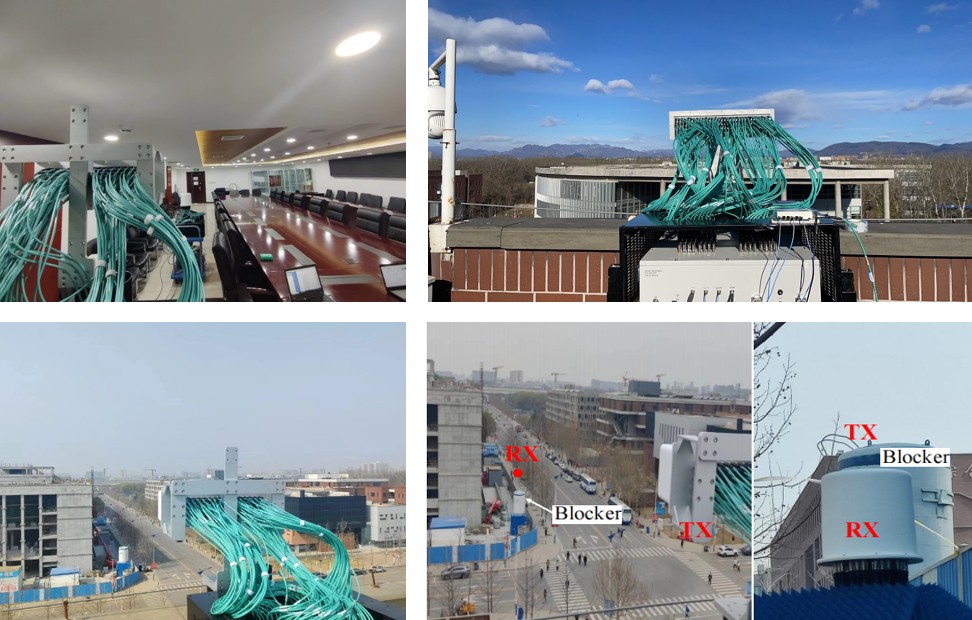}
\caption{XL-MIMO channel measurement environment and campaign.}
\label{scenario}
\end{figure}

To investigate the evolution of channel characteristics from conventional MIMO to XL-MIMO systems, channel measurements were conducted in both outdoor and indoor environments, as illustrated in Fig. \ref{scenario}. For example, The TX was equipped with a uniform planar array (UPA) comprising of 32$\times$2 dual-polarized antenna elements, that is, 32 elements in the horizontal and 2 in the vertical dimension. To emulate a larger array, a mechanical sliding platform was utilized to extend the array by four translations horizontally and three vertically, yielding a virtual XL-MIMO array with 128$\times$6 dual-polarized elements, totaling 1536 antenna elements.

\section{Channel Characteristics and Model}
\label{sec:IV}
This section focuses on the channel characteristics and modeling issues of XL-MIMO. Depending on the system architecture and the deployment method, XL-MIMO antenna can be classified into four types: co-located, cell-free, sparse, and movable. In this section, we will elaborate on the channel characteristics of each of these four types of XL-MIMO and provide corresponding channel models.

\subsection{Co-located XL-MIMO}

\subsubsection{RMS Angular Spread}
\par The root mean square (RMS) angular spreads represent the power of MPCs spreading over the angle, serving as a crucial  second-order statistic for characterizing the dispersion of the power angular profile and can be calculated as
\begin{equation}
\begin{split}
\psi_{rms}& = \sqrt{\dfrac{\sum_{l=1}^{L}(\psi_l-\psi_{mean})^2P(\psi_{l})}{\sum_{l=1}^{L}P(\psi_{l})}} ,
\end{split} 
\end{equation}
\\ where
$\psi_l$ denotes the azimuth angle $\phi_l$ or elevation angle $\theta_l$, and $\psi_{rms}$ denotes the RMS azimuth angular spread $\phi_{rms}$ or RMS elevation angular spread $\theta_{rms}$. As shown in Table \ref{angular_spread}, ASA, ESA, ASD, and ESD are angular spreads of the azimuth angle of arrival, the elevation angle of arrival, the azimuth angle of departure, and the elevation angle of departure, respectively \cite{Miao2024Measurement}.

\begin{table}[htbp]
	\centering
	\caption{The RMS angular spread in mid-band vs. 3GPP}
	\label{angular_spread}
	\renewcommand{\arraystretch}{1.5}
	\setlength{\tabcolsep}{3mm}
	\begin{tabular}{cc|c|c|c|c}
		\hline \hline
	 \multicolumn{2}{c|}{\multirow{2}{*}{\makecell{RMS \\Angular Spread}}}  &  \multicolumn{2}{c|}{Measurement}  &  \multicolumn{2}{c}{3GPP\cite{3gpp38.901}}\\
		\cline{3-6}
   	&	& \makecell{LOS } &\makecell{NLOS }
		&  \makecell{LOS } &\makecell{NLOS }  \\ \hline
		\multirow{2}{*}{\makecell{ASA}} &$\mu$&1.49&1.65 &1.81&1.87\\ 
		\cline{2-6}
		&$\sigma$ &0.18&0.16&0.20&0.11\\ 
		\hline
		\multirow{2}{*}{\makecell{ASD}} &$\mu$&0.85&1.01 &1.15&1.41\\ 
	\cline{2-6}
	&$\sigma$ &0.26&0.28&0.28&0.28\\ 
	\hline
		\multirow{2}{*}{\makecell{ESA}} &$\mu$&1.26&1.36 &0.95&0.26\\ 
	\cline{2-6}
	&$\sigma$ &0.19&0.13&0.16&0.16\\ 
	\hline
		\multirow{2}{*}{\makecell{ESD}} &$\mu$&1.06&0.96 &-&-\\ 
	\cline{2-6}
	&$\sigma$ &0.30&0.30&-&-\\ 
		\hline	\hline
	\end{tabular}
\end{table}

\par In Table \ref{angular_spread}, it is found that in the 6 GHz band, the mean value of ESD is lower than that of ESA, which aligns with the azimuth angular spread. This can be attributed to the positioning of the Tx antenna on the roof of the building,  where its height exceeds that of the surrounding scatterers, leading to fewer reflection and scattering paths. In contrast, the measurement route contains many scatterers with a height higher than that of the antenna at the Rx end, resulting in more reflection and scattering paths being received.

\subsubsection{Channel Hardening}

Channel hardening and favorable propagation are two fundamental pillars of massive MIMO theory. Channel hardening stabilizes the communication link for a single user by causing its overall channel gain to converge to a deterministic value, thereby eliminating the detrimental effects of small-scale fading. Favorable propagation, on the other hand, enables multi-user efficiency by ensuring that the channel vectors of different users become nearly orthogonal, which minimizes inter-user interference. These two phenomena are not independent but are intrinsically linked manifestations of a single underlying mathematical reality: the asymptotic behavior of the channel matrix's Gram matrix. Specifically, for an $M\times N$ channel matrix $\mathbf{H}$, where $M$ is the number of receiving antenna elements or users, the orthogonality of the normalized Gram matrix, expressed as 
\begin{equation}
    \frac{1}{N} \mathbf{H} \mathbf{H}^H \to \mathbf{I}_M .
\end{equation}

To quantitatively describe the channel orthogonality, the condition number is analyzed in \cite{tang_hardning}. The standard condition number is defined as the ratio of the largest to the smallest eigenvalue of $\mathbf{HH}^H$, 
which gives a measure of the relative conditioning (or rank deficiency) of a matrix.
The eigenvalue $\lambda$ can be obtained by singular value decomposition of the matrix; it is the square of the singular value. A large condition number indicates that the subchannels are strongly correlated, while a condition number of 1 indicates that the subchannels are completely orthogonal.

\begin{figure}[!htbp]
\setlength{\abovecaptionskip}{0cm}  
\centering
\includegraphics[width=0.45\textwidth]{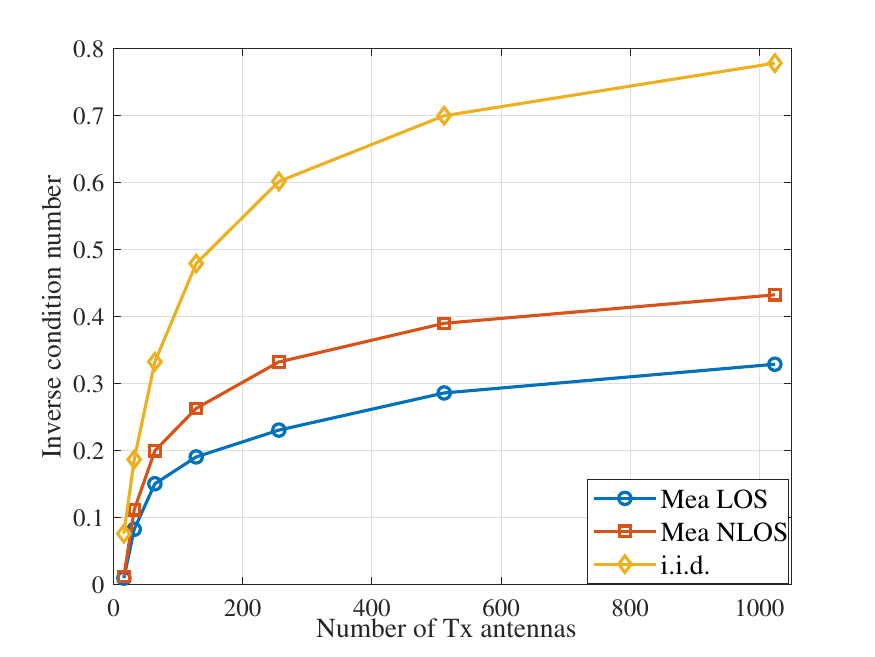}
\caption{Inverse condition number vs number of Tx antennas for measured and i.i.d. channel in the FR3 band.}
\label{invcond}
\end{figure}
When there are fewer antennas, the number of conditions obtained by measurement may be very large. To better display the results, we plot the inverse condition number so that the values are distributed between 0 and 1. As shown in Fig. \ref{invcond}, as the number of antennas increases, the inverse condition number gradually increases, indicating that the orthogonality of the channel is gradually stronger \cite{tang_hardning}. The channel hardening degree is the highest under the i.i.d. channel, and the channel orthogonality is also the strongest. The channel orthogonality is also the weakest in the LOS environment, with the lowest degree of channel hardening.

\subsubsection{Channel Capacity}

\par
\begin{figure}[htbp]
	\xdef\xfigwd{\columnwidth}
	\setlength{\abovecaptionskip}{0 cm}
	\centering
	\begin{tabular}{cc}
		\includegraphics[width=4.5cm,height=3.5cm]{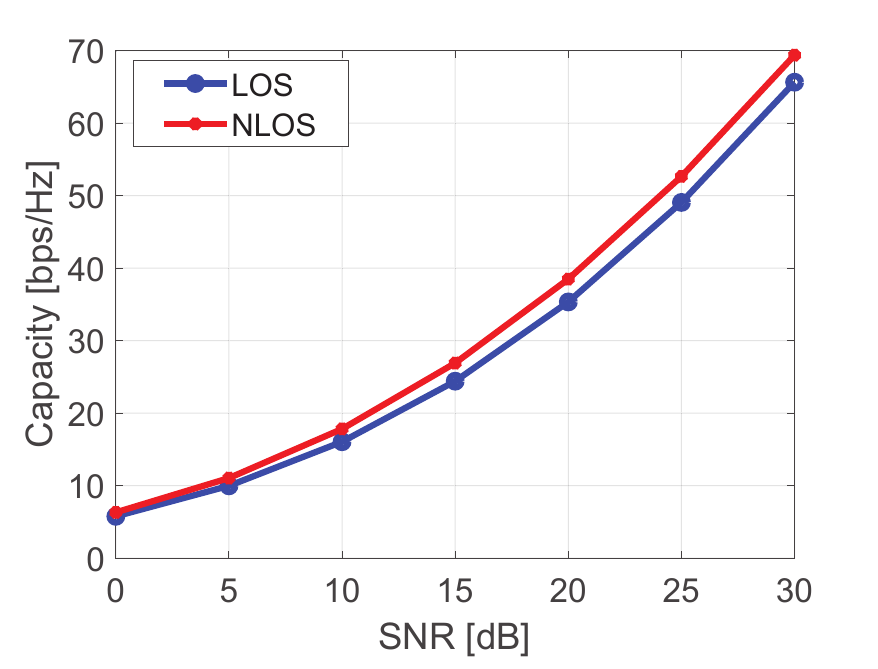} \hspace{-8mm}
		& \includegraphics[width=4.5cm,height=3.5cm]{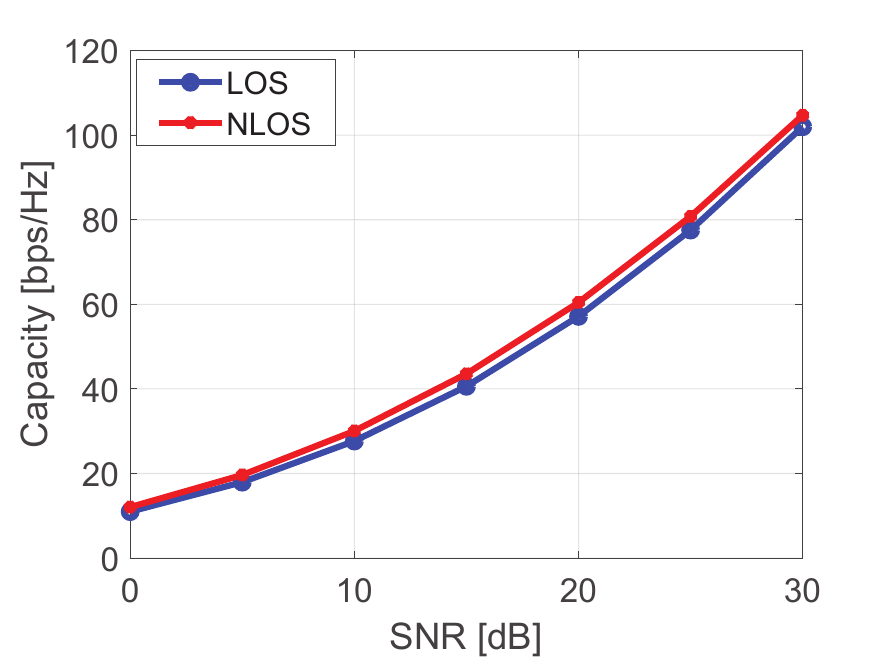}\\
		{\footnotesize\sf (a)} &	{\footnotesize\sf (b)} \\	
	\end{tabular}
	\caption{The channel capacity in the 6 GHz band. (a) 16 Rx antenna elements. (b) 56 Rx antenna elements.}
	\label{fig:Figure6_capacity}
\end{figure}

 Fig. \ref{fig:Figure6_capacity} gives the influence of the environments and number of elements on channel capacity at the 6 GHz band in the urban macrocell (UMa) scenario \cite{Miao2024Measurement}. In the NLOS environment, the channel capacity is generally higher compared to the LOS environment. This is primarily due to the increased complexity of the NLOS environment, where the multipath components exhibit greater degrees of freedom in the angle domain. Additionally, as the number of antenna array elements increases, the channel capacity shows a significant improvement, as illustrated in Fig. \ref{fig:Figure6_capacity}. In particular, with a higher number of antenna elements, the channel capacity grows more rapidly with an increasing signal-to-noise ratio (SNR).

\par These findings highlight that increasing the number of antenna elements is one of the most effective methods to enhance channel capacity. This insight is a key reason why XL-MIMO is emerging as a potential cornerstone technology for 6G communication systems. XL-MIMO leverages much larger-scale antenna arrays, and with further expansion of the antenna scale, it promises to deliver even higher throughput and spectral efficiency, making it a critical enabler for future wireless communication systems.

\begin{figure}[!htbp]
\centering
\includegraphics[width=1.0\linewidth]{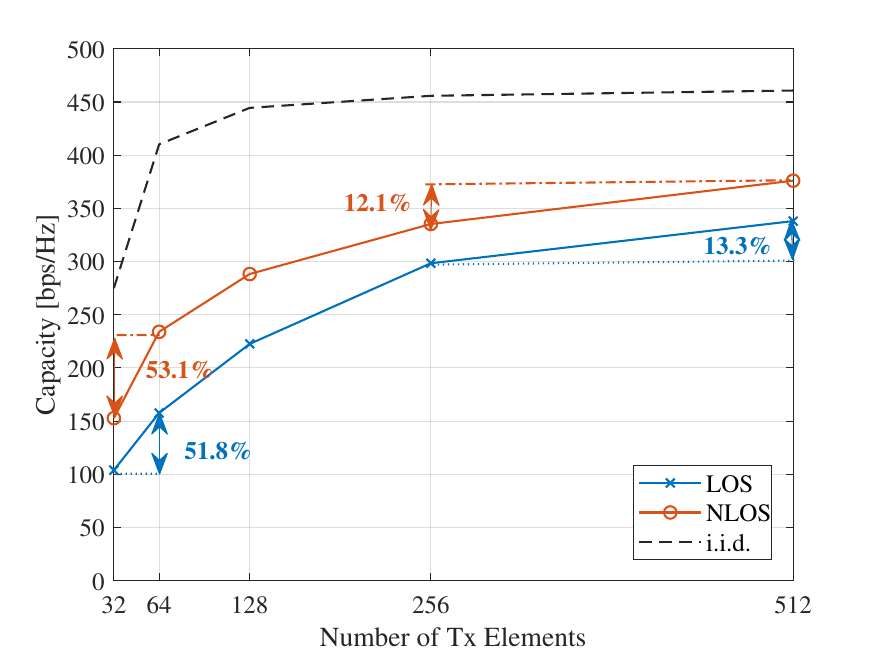}
\caption{Channel capacity for various number of elements, SNR=25 dB.}
\label{Capcompare}
\end{figure}

In order to compare the indoor environment, we also observe the capacity characteristics of the corridor scenario\cite{Wei2024Corridor}. Fig. \ref{Capcompare} illustrates how the channel capacity varies with the number of transmitting antenna elements increasing from 32 to 512. The results indicate that the growth rate of channel capacity decreases as the number of transmitting antenna elements is sequentially doubled, from 51.8$\%$ to 13.3$\%$ for LOS and from 53.1$\%$ to 12.1$\%$ for NLOS. The channel capacity under i.i.d. channel does not show a substantial increase  when the number of transmitting antenna elements exceeds 64. This is due to the limited number of 56 antenna elements at the receiver end, causing the capacity to be saturated when the number of transmitting antenna elements surpasses this threshold.

Furthermore, there exists a relatively significant disparity between the channel capacity under the corridor scenario and that under the i.i.d. channel. Specifically, the channel capacity in the LOS and NLOS environments achieves a maximum of 73.4$\%$ and 81.6$\%$ of the i.i.d. channel capacity, respectively. This difference arises due to the relatively closed space of the corridor scenario, where the spatial freedom of the MPCs is limited. The research presented in Fig. \ref{Capcompare} indicates that XL-MIMO performs comparably  to the i.i.d. channel in open environments with relatively abundant scattering. 

\subsubsection{Spherical-Wave Property}
\par
\begin{figure}[!htbp]
    \centering
    \begin{subfigure}[t]{0.45\textwidth} 
        \includegraphics[width=\linewidth]{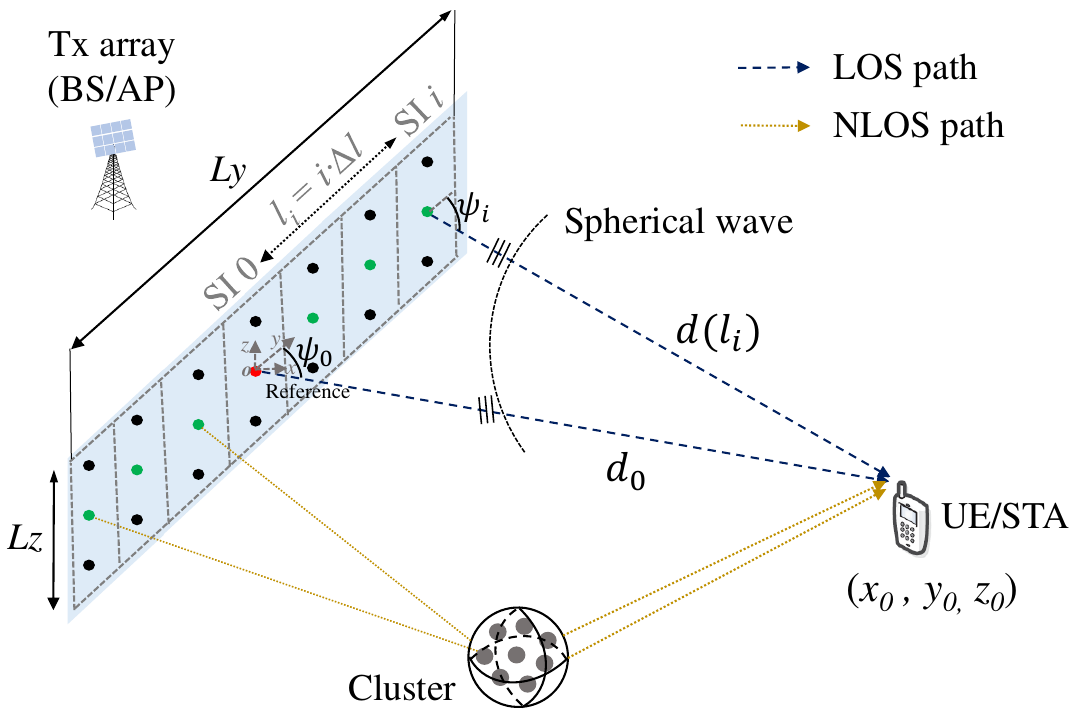}
        \caption{}
        \label{figure_capacityNTx_LOS}
    \end{subfigure}
    \hfill 
    \begin{subfigure}[t]{0.45\textwidth}
        \includegraphics[width=\linewidth]{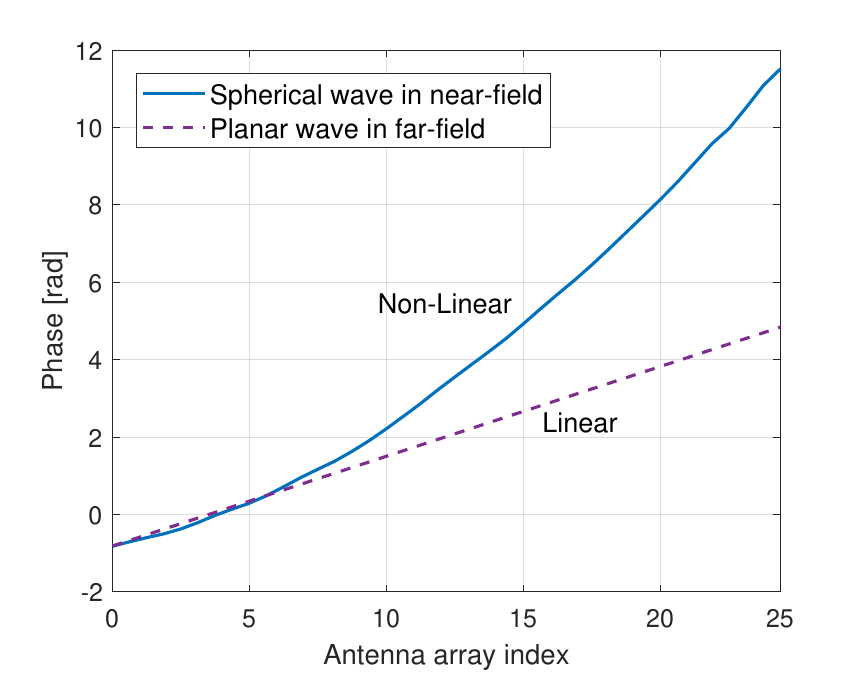}
        \caption{}
        \label{figure_capacityNTx_NLOS}
    \end{subfigure}
    
    \caption{The phase verification in the new mid-band. (a) Near-field multipath channel propagation model. (b) The wavefront model from far-field to near-field. }
    \label{fig:Nearfield_wave}
\end{figure}

 When the array antennas reach such a large dimension, spatial non-stationary characteristics are more likely to appear along the array. The different parts on the array may have different views of the propagation environment and paths. In the near-field region, the phase and difference in traveling distance at antennas are different from far-field propagation. Based on the sliding window measurement method, the antenna array is divided into different subarrays, and the subarrays are stationary channels. The subarray is defined as the stationary interval. Considering that the angle model related to the subset of array elements needs to be established in the array domain, we found that the angle can be introduced to model the spherical-wave signal \cite{miao2025far}. Fig. \ref{fig:Nearfield_wave} shows the verification of the spherical-wave propagation property.

\par As shown in Fig. \ref{fig:Nearfield_wave}(a), the Tx side is a large-scale MIMO array and we take the first-bounce cluster or user equipment (UE) on the downlink as an example. The near-field channel propagation model is characterized by introducing distance parameters. In Fig. \ref{fig:Nearfield_wave}(b), the phase derived from the simulation aligns closely with the established signal model\cite{miao_near}. By applying correlation to compare the two sets of values, it is evident that the correlation between them is remarkably strong, suggesting that the theoretical model exhibits excellent accuracy.

\subsubsection{Stationary Interval Division}

To accurately model the SnS phenomenon, the work \cite{zwr1} proposes a method for stationary sub-interval partitioning based on channel characteristics. This approach assumes that the channel remains stationary within each sub-interval, but exhibits non-stationary across different intervals. The method comprehensively considers factors such as channel correlation, delay spread (DS), azimuth angle spread of departure (ASD), and the birth-death behavior of MPCs for sub-interval partitioning. By analyzing the independence of sub-intervals, this study demonstrates that the proposed method outperforms the conventional averaging technique in terms of sub-interval partitioning.

\par For ULA, we can first obtain the MPCs parameter data $X$, $X = \{x_{1}, x_{2}, \ldots, x_{K}\}$ based on RT simulation, where $x_{1}$ represents the MPCs parameters of the first array element, including power, delay, phase, and angle. Then we can calculate the channel impulse response (CIR) for each transmitting element, which is used for subsequent characteristics analysis.

After processing the parameters\cite{zwr1} , we can proceed with the partitioning of stationary intervals based on the characteristic analysis in Section III. Initially, we preselect interval nodes based on the birth-death rate of the MPCs, with the selection method outlined by the following expression
\begin{align}
S(k) = \begin{cases}
    1 & \text{if } p_{k}^{n} - p_{k-1}^{n} \ge 3\, \text{dB} \\
   -1 & \text{if } p_{k}^{n} - p_{k-1}^{n} \le -3\, \text{dB} \\
    0 & \text{else}\\
\end{cases},
\end{align}
where $k$ belongs to the range 1 to $K$, $K$ is the number of transmitting array elements, $n$ belongs to the range 1 to $N$, $N$ is the total number of MPCs from the $k$-th array element. The value of $S(k)$ represents the birth and death situation at the $k$-th element, $S(k) = 1$ indicates the birth of a new MPC at the $k$-th element, $S(k) = -1$ indicates the death of a MPC at this location, and $S(k) = 0$ indicates no birth or death phenomenon at the $k$-th element. The specific selection criterion is based on the power variation of multipath signals on the array domain.
$p_k^n$ represents the power of the $n$-th path at the $k$-th array element, and $p_{k-1}^n$ represents the power of the $n$ path at the $(k-1)$-th array element. If there exists a path for which $p_k^n-p_{k-1}^n \ge$ 3 dB, then set $S(k)=1$, when $p_{k}^n-p_{k-1}^n \le$ 3 dB, set $S(k)=-1$. By iterating through all elements, we obtain the complete $S$ vector.

After obtaining the $S$ vector, further refinement of the selection of interval nodes is necessary, utilizing channel correlation, DS, and ASD. By analyzing the variations in the parameters of the array elements, elements from $S$ are identified as potential interval boundaries. 
First, by analyzing the changes in correlation, if a certain interval can be considered stationary, the correlation coefficients for sub-channels within that interval will be relatively large and exhibit minimal fluctuations. When a new element is introduced to this interval, a significant increase in fluctuations may indicate that the newly added element serves as a boundary. The fluctuations are calculated using the Mean Absolute Deviation (MAD).
\begin{align}
M A D=\frac{1}{n} \sum_{i=1}^{n}\left|x_{i}-\bar{x}\right|,
\end{align}
where $n$ is the number of the data, $x_{i}$ represents the $i$-th data, and $\bar{x}$ is the mean of all data.
A similar approach can be applied to DS and ASD. If adding an element to an interval leads to a significant  rise in parameter fluctuations, then that element may also be a boundary element. Together considering channel correlation, DS, and ASD, the following equation is utilized for evaluation.
\begin{align}
w_{c} \cdot \frac{M_{k}^{C}}{M_{k-1}^{C}}+w_{a} \cdot \frac{M_{k}^{A}}{M_{k-1}^{A}}+w_{d} \cdot \frac{M_{k}^{D}}{M_{k-1}^{D}}>\rho ,\label{eq:lianhe}
\end{align}
where $M_{k}^{C}$ represents the fluctuation of channel correlation within the interval when the $k$-th element is chosen as the boundary, and $M_{k-1}^{C}$ represents the fluctuation when the $(k-1)$-th element is taken as the boundary. 
Similarly, $M_{k}^{A`}$ and $M_{k}^{D}$ measure the fluctuation of ASD and DS, respectively. 
$w_{c}$, $w_{a}$ and $w_{d}$ represent the weights assigned to the fluctuation values of correlation coefficient, ASD, and DS, respectively. 
$\rho$ is the threshold for judgment. 
The inequality \eqref{eq:lianhe} is used to evaluate each element in $S$, and if the calculated result at a certain element exceeds the threshold, that element is considered a boundary node. After completing the traversal, the final interval partitioning result $Z$ is obtained.

\begin{figure}[htbp]
\centerline{\includegraphics [width=0.45\textwidth] {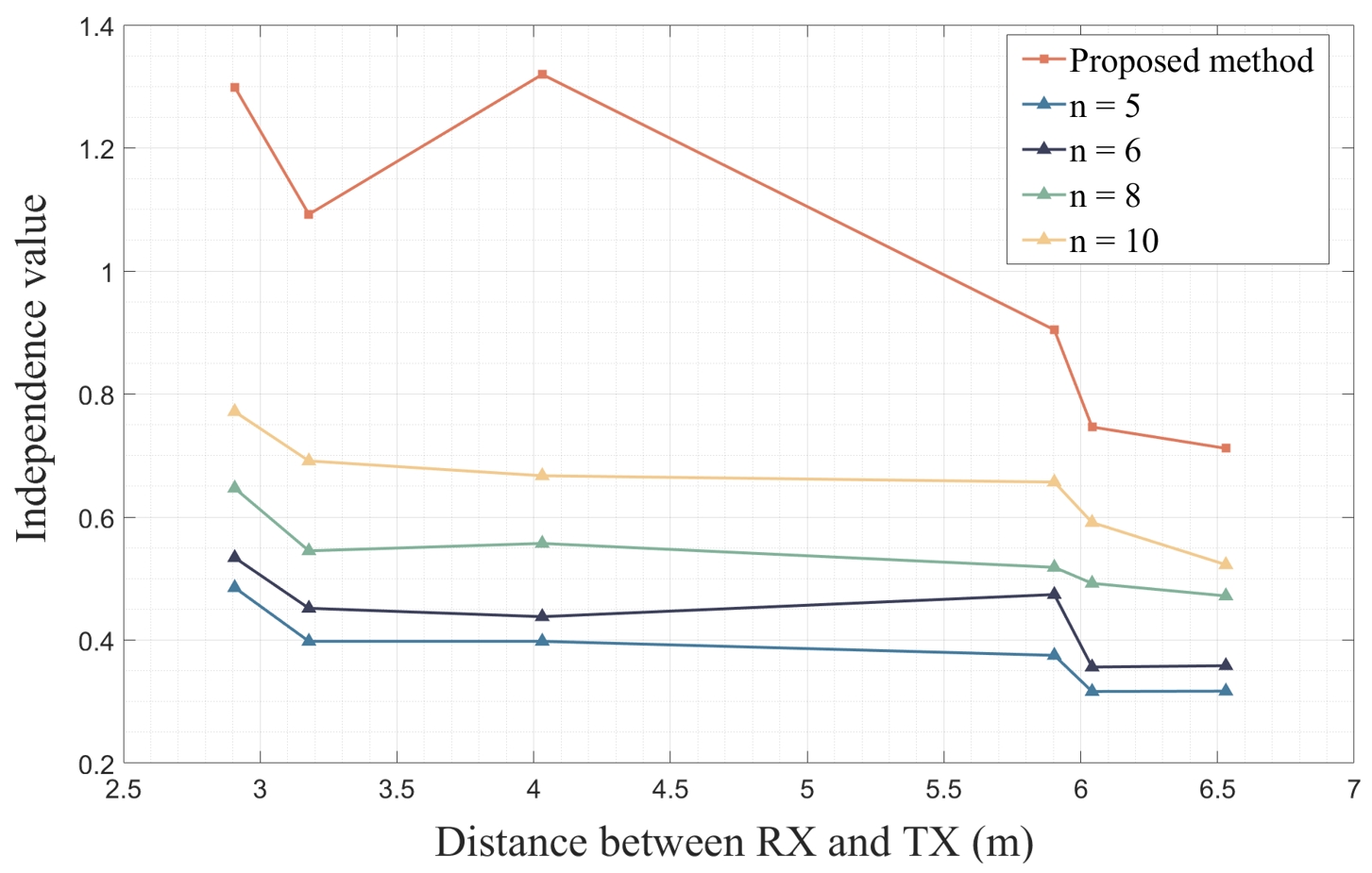}}
\caption{Degree of independence of sub-intervals obtained by different partitioning methods. The $n$ represents the number of array elements in each sub-interval during average partitioning.}
\label{fig:duli}
\end{figure}

After completing the partitioning, the reconstructed CIR for each sub-interval can be revived. Then applying  equation \eqref{eq:mubiao}, we analyzed the independence of the sub-intervals obtained based on channel characteristics and average partitioning. The results are presented in Fig.~\ref{fig:duli}.
\begin{align}
D =\frac{1}{N \cdot (M-1)} \sum_{k=1}^{M-1} \sum_{i=1}^{N}\left|h_{k+1}(i)-h_{k}(i)\right|,\label{eq:mubiao}
\end{align}
where $N$ and $M$ respectively represent the number of delay bins and sub-intervals, $h_{k}$ represents the reconstruction CIR of the $k$-th sub-interval.

In Fig.~\ref{fig:duli}, the horizontal and vertical axes correspond to the Tx-Rx distance and independence level, respectively. It is evident that since traditional partitioning techniques neglect environmental factors, the independence of sub-intervals is influenced by the size of the intervals, thereby resulting in poor robustness. In contrast, when partitioning is based on channel characteristics, the method accurately assigns elements with different scattering environments to distinct intervals. This results in enhanced accuracy and robustness in the obtained outcomes.

Additionally, it is observed that as the distance between the transmitter and receiver increases, the independence of the sub-intervals for different methods decreases. This occurs because with the distance surpassing the Rayleigh distance, the scattering environments in different sub-intervals become identical, the SnS phenomenon becomes less pronounced, and the independence of the sub-intervals becomes zero.

\subsubsection{Channel Model for 3GPP}
The modeling of small-scale fading in the 5G era was largely constrained to far-field and spatial stationary conditions, an approach that inherently depended on the assumptions of planar wavefronts and constant path parameters over the antenna array. In this model, a cluster is conceptually represented as a scattering region comprising $M$ rays, with a total of $N$ clusters assumed. The system comprises $S$ transmit antennas and $U$ receive antennas.

However, as antenna array apertures expand for 6G, channel models must evolve to capture near-field (NF) propagation and spatial non-stationarity. In contrast to conventional far-field assumptions, 6G channel models capture element-dependent ray characteristics by allowing parameters such as azimuth/elevation angles of departure (AOD/ZOD) and arrival (AOA/ZOA) to change along the antenna array, thereby reflecting the different propagation geometries perceived by spatially separated elements. Near-field effects are modeled by introducing spherical-wave source distances (e.g., $d_1$ and $d_2$), which define the effective wavefront curvature. Consequently, phase evolution across the array is governed by true geometric distances rather than planar approximations. To incorporate SnS, the framework introduces element-wise power weighting factors applied to each ray. This mechanism mathematically captures physical phenomena such as partial blockage and incomplete scattering across the massive base station aperture, as well as element-dependent self-blockage (e.g., hand/head grip effects) at the user equipment. Furthermore, to enhance stochastic realism, the model integrates two novel variability features: cluster number variability, where the number of active clusters is drawn from a scenario-dependent distribution rather than being fixed; and polarization power variability, which applies ray-specific random weights to cross-polarization coefficients. These extensions constitute a unified 6G small-scale fading model capable of consistently reproducing spherical-wave propagation, spatial non-stationarity, and realistic polarization dynamics \cite{near_field3g}.

\begin{table}[!h]
\renewcommand\arraystretch{1}
 \caption{Definitions of key channel model parameters}
 \centering
 \begin{tabular}{m{3.8cm}<{\centering}|m{4cm}<{\centering}}
 \hline \hline
 Model Parameters & Definitions \\
\hline
\(\phi_{{\rm{LOS},AOD,}u,s}\) \(\phi_{{\rm{LOS},AOA,}u,s}\) \(\theta_{{\rm{LOS},ZOD,}u,s}\) \(\theta_{{\rm{LOS,ZOA}},u,s}\) & Near-field: AOD, AOA, ZOD, and ZOA of the Tx-Rx \((s, u)\) pair\\
\hline
\(\phi_{n,m,{\rm{AOD}},s}\), \(\phi_{n,m,{\rm{AOA}},u}\), \(\theta_{n,m,{\rm{ZOD}},s}\), \(\theta_{n,m,{\rm{ZOA}},u}\) &{Near-field: AOD, ZOD (Tx element \(s\)) and AOA, ZOA (Rx element \(u\)) for ray \(m\) in cluster \(n\)}\\
 \hline
 \raisebox{-1ex}{\({{{\vec r}_{u,s}}}\)} & { Near-field: Vector from Tx element \(s\) to Rx element \(u\)} \\
 \hline
\(d_{1,n,m}\) & Distance from BS to the spherical-wave source for ray \(m\) in cluster \(n\) \\
 \hline
\(d_{2,n,m}\) & Distance from UE to the spherical-wave source for ray \(m\) in cluster \(n\) \\
 \hline
\(\alpha_{s,n,m}\) & Power attenuation factor for ray \(m\) in cluster \(n\) at Tx antenna element \(s\)\\
 \hline
 \(\beta_{u}\) & Power attenuation factor at Rx antenna element \(u\) \\
 \hline
 {\({F_{rx,u,\theta }}\), \({F_{rx,u,\phi }}\) }&Field patterns of Rx element \(u\) along spherical basis vectors \(\hat{\theta}\) and \(\hat{\phi}\), respectively \\
\hline
{\({F_{tx,s,\theta }}\), \({F_{tx,s,\phi }}\) }& Field patterns of Tx element \(s\) along spherical basis vectors \(\hat{\theta}\) and \(\hat{\phi}\), respectively \\
\hline
{\({\hat{r}_{rx,n,m }}\), \({\hat{r}_{tx,n,m }}\)}& Spherical unit vectors of ray \(m\) in cluster \(n\) at the Rx and Tx, respectively \\
 \hline
{\({\bar{d}_{rx,u}}\), \({\bar{d}_{tx,s}}\) }& Location vectors of Rx element \(u\) and Tx element \(s\), respectively\\
 \hline
{\({{d}_{\rm{3D}}}\) }& 3D distance between the reference points of the BS and the UE \\
 \hline \hline
\end{tabular}
 \label{Tab_Par}
\end{table}

The channel impulse response, denoted as $H_{u,s}(\tau, t)$ between the $u$-th Rx and $s$-th Tx antenna element, is modeled as a superposition of LOS and NLOS components. As formulated in (\ref{equ_CIR_LOS}), these components are scaled by the Ricean K-factor:
\begin{equation}
\begin{aligned}
H_{u,s}^{{\rm{LOS}}}\left( {\tau ,t} \right) =& \sqrt {\frac{1}{{{K_R} + 1}}} H_{u,s}^{{\rm{NLOS}}}\left( {\tau ,t} \right) + \\
&\sqrt {\frac{{{K_R}}}{{{K_R} + 1}}} H_{u,s,1}^{{\rm{LOS}}}\left( t \right)\delta \left( {\tau - {\tau _1}} \right),
\end{aligned}
\label{equ_CIR_LOS}
\end{equation}
where \(K_R\) is the linear Ricean K-factor (with $K_R = 0$ indicating purely NLOS conditions). The NLOS component consists of \(N\) clusters, each containing \(M\) rays, and is expressed as:
\begin{equation}
\begin{aligned}
H_{u,s}^{{\rm{NLOS}}}\left( {\tau ,t} \right) = &\sum\limits_{n = 1}^2 {\sum\limits_{i = 1}^3 {\sum\limits_{m \in {R_i}} {H_{u,s,n,m}^{{\rm{NLOS}}}(t)\delta \left( {\tau - {\tau _{n,i}}} \right) } } } \\
&+ \sum\limits_{n = 3}^N {\sum\limits_{m = 1}^M {H_{u,s,n,m}^{{\rm{NLOS}}}(t)\delta \left( {\tau - {\tau _n}} \right)}},
\end{aligned}
\label{equ_CIR_NLOS}
\end{equation}
where rays in the two strongest clusters ($n=1,2$) are spread in delay into three sub-clusters indexed by $i$. 

To accurately capture 6G channel propagation features, the explicit channel coefficients for the LOS component, $H_{u,s,1}^{\mathrm{LOS}}(t)$, and the NLOS rays, $H_{u,s,n,m}^{\mathrm{NLOS}}(t)$, are mathematically expanded in (\ref{equ_H_LOS_6G}) and (\ref{equ_H_NLOS_6G}), respectively. It is worth noting that the proposed channel model incorporates two improvements:
\begin{itemize}
    \item Spherical Phase: The phase terms are determined by the precise geometric distances ($d_{1,n,m}$, $d_{2,n,m}$) between antenna elements and spherical-wave sources, replacing the linear phase approximation of far-field models. 
    \item SnS Power: The terms $\beta_{u}^{\rm{SnS}}$ and $\alpha_{s,n,m}^{\rm{SnS}}$ represent the power attenuation on the UE and BS sides, respectively, allowing element-wise power variations.
\end{itemize}

The coefficients involve the polarization matrix $\mathbf{\Phi}_{n,m}$, given by 
\begin{equation}
\begin{split}
\mathbf{\Phi}_{n,m} =
\begin{bmatrix}
\sqrt{\eta_{n,m,\theta \theta}} e^{j \Phi_{n,m}^{\theta\theta}} & \sqrt{\eta_{n,m,\theta \phi}\kappa_{n,m}^{-1}} e^{j \Phi_{n,m}^{\theta\phi}} \\
\sqrt{\eta_{n,m,\phi \theta}\kappa_{n,m}^{-1}} e^{j \Phi_{n,m}^{\phi\theta}} & \sqrt{\eta_{n,m,\phi \phi}}e^{j \Phi_{n,m}^{\phi\phi}}
\end{bmatrix}
\end{split}
\end{equation}
where $\kappa_{n,m}$ denotes the cross-polarization power ratio (XPR), and $\Phi$ represents independent random phases uniformly distributed over $[0, 2\pi)$. 

For clarity, Table~\ref{Tab_Par} provides a comprehensive definition of the key geometric and physical parameters used in these equations, including the spherical unit vectors, field patterns, and element-wise spherical-wave distance variables. Finally, large-scale fading effects (path loss and shadowing) are applied to these normalized small-scale fading coefficients.

\begin{figure*}
\begin{equation}
\begin{aligned}
H_{u,s,1}^{\rm{LOS}}&\left( t \right) = \sqrt{\beta_{u}^{\rm{SnS}} \alpha_{s,1}^{\rm{SnS}}} \left[ {\begin{array}{*{20}{c}}
{{F_{rx,u,\theta }}\left( {{\theta _{{\rm{LOS,ZOA}},u,s}},{\phi _{{\rm{LOS,AOA}},u,s}}} \right)}\\
{{F_{rx,u,\phi }}\left( {{\theta _{{\rm{LOS,ZOA}},u,s}},{\phi _{{\rm{LOS,AOA}},u,s}}} \right)}
\end{array}} \right]\left[ {\begin{array}{*{20}{c}}
1&0\\
0&{ - 1}
\end{array}} \right] \\
&\cdot \left[ {\begin{array}{*{20}{c}}
{{F_{tx,s,\theta }}\left( {{\theta _{{\rm{LOS,ZOD}},u,s}},{\phi _{{\rm{LOS,AOD}},u,s}}} \right)}\\
{{F_{tx,s,\phi }}\left( {{\theta _{{\rm{LOS,ZOD}},u,s}},{\phi _{{\rm{LOS,AOD}},u,s}}} \right)}
\end{array}} \right] \exp \left( { - j2\pi \frac{{{d_{{\rm{3D}}}}}}{{{\lambda _0}}}} \right)\exp \left( { - j2\pi \frac{{\left| {{{\vec r}_{u,s}}} \right| - {d_{{\rm{3D}}}}}}{{{\lambda _0}}}} \right)\exp \left( {j2\pi \frac{{\hat r_{rx,{\rm{LOS}}}^T\bar v}}{{{\lambda _0}}}t} \right).
\end{aligned}
\label{equ_H_LOS_6G}
\end{equation}
\end{figure*}

\begin{figure*}
\begin{equation}
\begin{aligned}
H&_{u,s,n,m}^{\rm{NLOS}}\left( t \right) = \sqrt{\beta_{u}^{\rm{SnS}} \alpha_{s,n,m}^{\rm{SnS}}} \sqrt {\frac{{{P_n}}}{M}} \left[ {\begin{array}{*{20}{c}}
{{F_{rx,u,\theta }}\left( {{\theta _{n,m,{\rm{ZOA}},u}},{\phi _{n,m,{\rm{AOA}},u}}} \right)}\\
{{F_{rx,u,\phi }}\left( {{\theta _{n,m,{\rm{ZOA}},u}},{\phi _{n,m,{\rm{AOA}},u}}} \right)}
\end{array}} \right]\mathbf{\Phi}_{n,m} \left[ {\begin{array}{*{20}{c}}
{{F_{tx,s,\theta }}\left( {{\theta _{n,m,{\rm{ZOD}},s}},{\phi _{n,m,{\rm{AOD}},s}}} \right)}\\
{{F_{tx,s,\phi }}\left( {{\theta _{n,m,{\rm{ZOD}},s}},{\phi _{n,m,{\rm{AOD}},s}}} \right)}
\end{array}} \right] \\
& \cdot\exp \left( {j2\pi \frac{{{d_{2,n,m}} - \left\| {{d_{2,n,m}} \cdot {{\hat r}_{rx,n,m}} - {{\bar d}_{rx,u}}} \right\|}}{{{\lambda _0}}}} \right)\exp \left( {j2\pi \frac{{{d_{1,n,m}} - \left\| {{d_{1,n,m}} \cdot {{\hat r}_{tx,n,m}} - {{\bar d}_{tx,s}}} \right\|}}{{{\lambda _0}}}} \right)\exp \left( {j2\pi \frac{{\hat r_{rx,n,m}^T\bar v}}{{{\lambda _0}}}t} \right).
\end{aligned}
\label{equ_H_NLOS_6G}
\end{equation}
\end{figure*}

\subsection{Cell-Free XL-MIMO}

Cell-free XL-MIMO is a distributed massive MIMO architecture in which multiple access points (APs) jointly and coherently serve all users over the same time–frequency resources, without cell boundaries\cite{Ngo2017cell-free}. By combining the advantages of distributed antenna systems (DAS) and MIMO, it represents a promising paradigm that offers improved coverage and performance\cite{interdonato2019ubiquitous}. 

\subsubsection{Channel Capacity}

The channel capacity characterizes the maximum information-carrying capacity of the channel, calculated as\cite{paulraj2003introduction}: 
\begin{equation}
\begin{split}
C=\frac{1}{B} \int_B \log _2 \operatorname{det}\left(\mathbf{I}_M+\frac{\rho}{N} H(f)H^H(f)\right) d f,
\end{split} 
\end{equation}
where  $B$ is the system bandwidth, $\mathbf{I}_M$ is an $M\times M$ unit matrix, $M$ is the number of Rx antenna elements, $\rho$ is the transmission SNR, $N$ is the number of Tx antenna elements and $ (·) ^H $ is a Hermitian operation. 

When the channel state information (CSI) is unknown at the BS side, allocating equal power across all antennas is the optimal strategy for capacity calculation \cite{goldsmith2005wireless,telatar1999capacity}. The mean channel capacity then is calculated by
\begin{equation}
\begin{split}
\tilde{C}=\frac{1}{K} \sum_{k=1}^K \log _2 \operatorname{det}\left(\mathbf{I}_M+\frac{\rho}{\beta N} \mathbf{H}_r \mathbf{H}_r^H\right),
\end{split} 
\end{equation}
where $\mathbf{H}_r$ is the discrete channel realization, $K$ is the total number of such realizations. A normalization factor $\beta $ is applied to ensure uniform average channel power gain across all $\mathbf{H}_r$ in each snapshot, which can be expressed as
\begin{equation}
\begin{split}
E\left\{\frac{1}{\beta}\left\|\mathbf{H}_r\right\|_F^2\right\}=N \cdot M,
\end{split} 
\end{equation}
where $\left\|·\right\|_F$ is the Frobenius norm.

The detailed measurement setup is described in \cite{zhen2025cellfreeversusconventionalmassive}. We first analyze how the channel capacity changes as the Rx moves along the measurement routes. Considering that the channel capacity differs markedly under LOS and NLOS conditions, the two cases are considered separately. The SNR of 25 dB selected is a commonly encountered value.
\begin{figure}[h]
    \centering
    \begin{subfigure}[t]{0.45\textwidth} 

        \includegraphics[width=\linewidth, clip,trim=50 60 20 60]{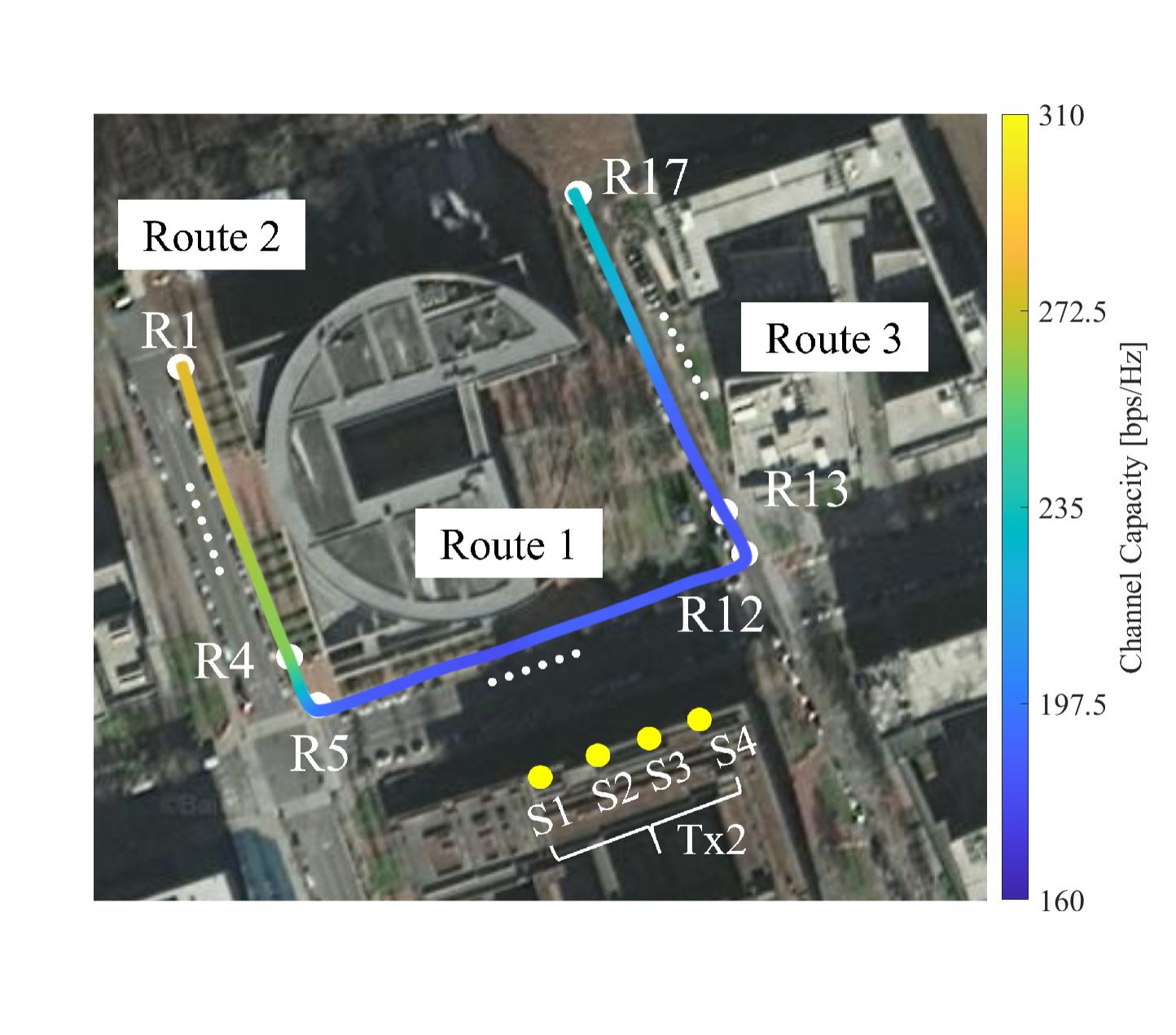}
        \caption{}
        \label{figure_capacityPoint_CF-mMIMO}
    \end{subfigure}
    \begin{subfigure}[t]{0.45\textwidth}
        \includegraphics[width=\linewidth, clip,trim=50 40 20 50]{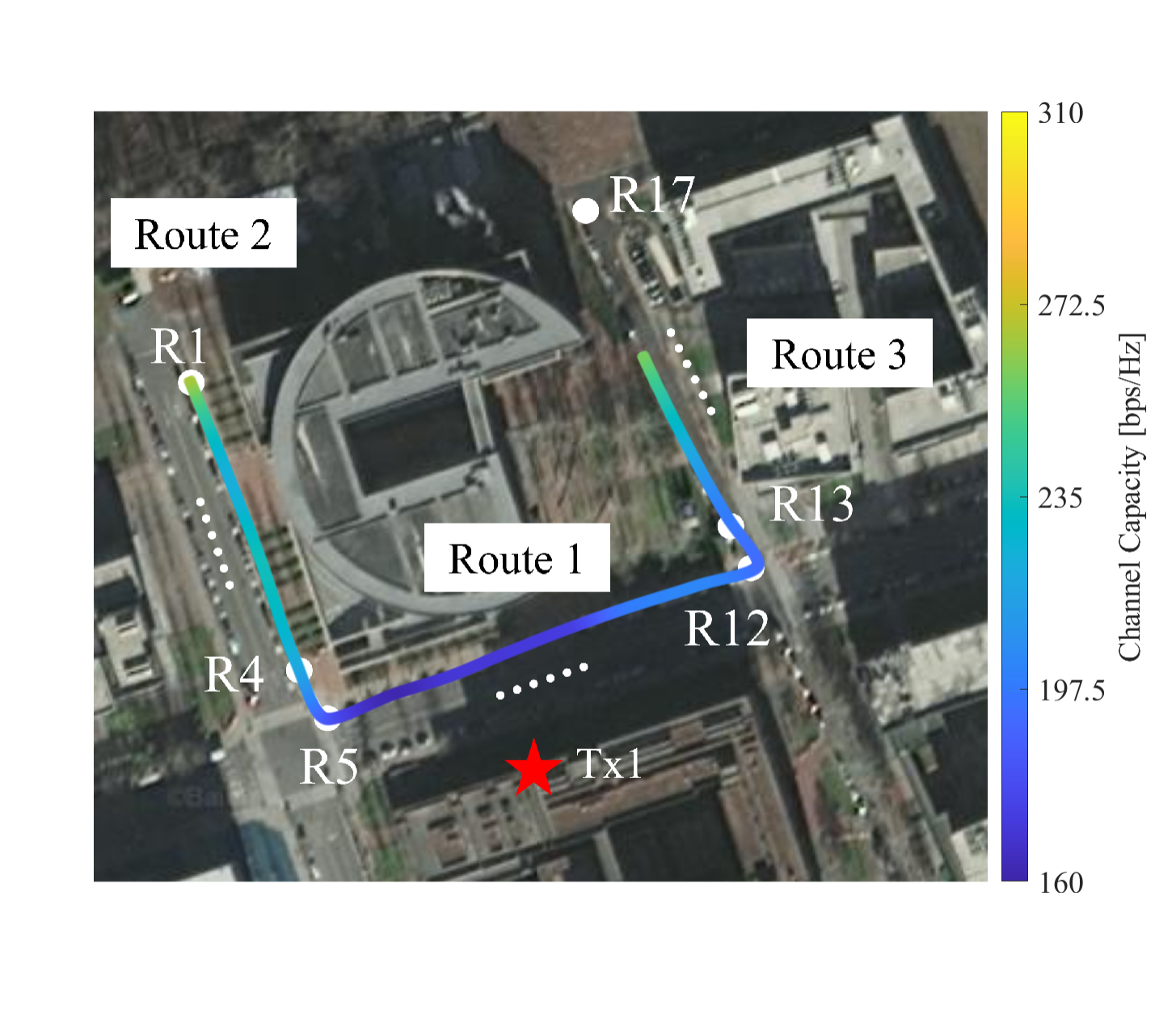}
        \caption{}
        \label{figure_capacityPoint_Conventional_mMIMO}
    \end{subfigure}
    
    \caption{The variation of channel capacity along the routes in the 15 GHz band. (a) CF-mMIMO. (b) Conventional mMIMO. }
    \label{figure_capacityPoint_SNR}
\end{figure}
Fig. \ref{figure_capacityPoint_SNR} presents heatmaps of the channel capacity variations as the Rx moves along the measurement routes, comparing the CF-mMIMO and Conventional configurations. In Fig.\ref{figure_capacityPoint_SNR}(a), the four yellow circles denote Subarrays 1 to 4, which together form a virtual large-scale antenna array referred to as Tx2. Fig.\ref{figure_capacityPoint_SNR}(b) illustrates the Conventional mMIMO configuration, where the red pentagrams represent the virtually co-located large-scale antenna array, designated as Tx1. 

As illustrated in Fig. \ref{figure_capacityPoint_SNR}, at Route 2 (NLOS condition) the channel capacity of CF-mMIMO exhibits a pronounced gain of more than 35 bps/Hz compared with Conventional MIMO, whereas the enhancement at points 5-8 (LOS condition) is not very evident. This is because the four sub-arrays for CF-mMIMO, which are geographically separated, form an exceptionally large virtual antenna array. Under NLOS condition, the presence of more incomplete scatterers becomes significant for such a large-scale Tx antenna array. This, in turn, intensifies the NLOS propagation environment, leading to more dispersed multipath components (MPCs) and a substantial increase in channel capacity.

\begin{figure}[h]
    \centering
    \begin{subfigure}[t]{0.45\textwidth} 
        \includegraphics[width=\linewidth]{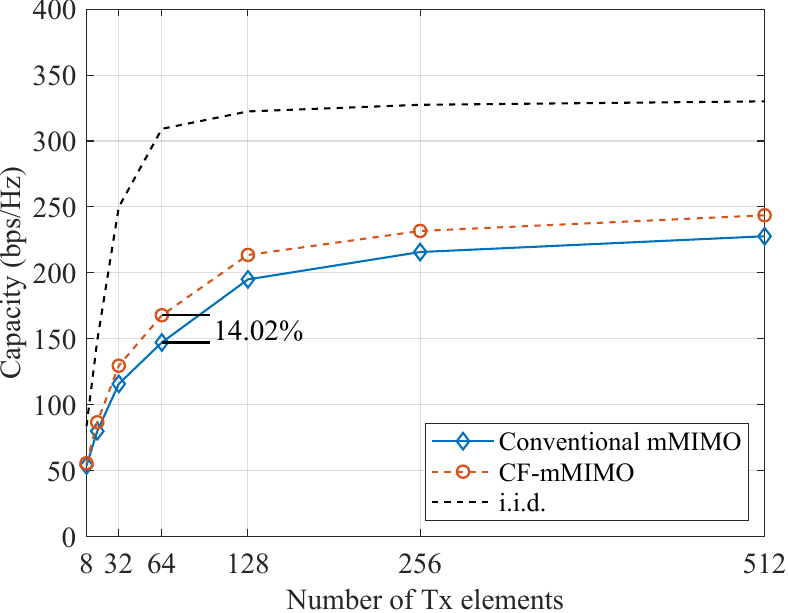}
        \caption{}
        \label{figure_capacityNTx_LOS}
    \end{subfigure}
    \hfill 
    \begin{subfigure}[t]{0.45\textwidth}
        \includegraphics[width=\linewidth]{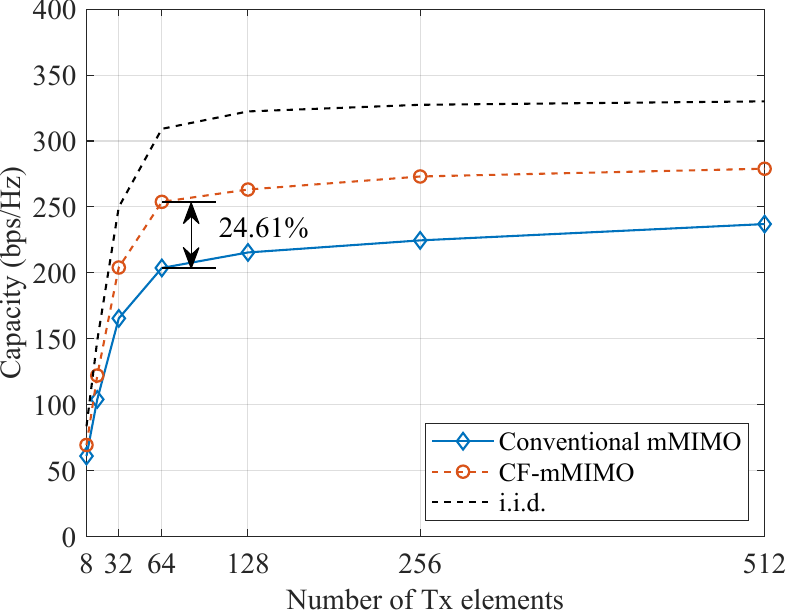}
        \caption{}
        \label{figure_capacityNTx_NLOS}
    \end{subfigure}
    
    \caption{The channel capacity of CF-mMIMO and Conventional mMIMO  under different numbers of Tx elements in the 15 GHz band. (a) LOS. (b) NLOS. }
    \label{figure_capacity_NTx}
\end{figure}

Besides the analyses given above, we then proceed to examine how the number of Tx elements influences the advantage in channel capacity for CF-mMIMO over the Conventional configuration. Fig. \ref{figure_capacity_NTx} illustrates the channel capacity of CF-mMIMO and Conventional mMIMO under both LOS and NLOS conditions, for varying numbers of antenna elements in the entire Tx array. The capacity of the Rayleigh channel is also illustrated, denoted by i.i.d. Note that SNR is 25 dB. 

It can be observed that channel capacity increases with the growth in the number of Tx elements under identical propagation conditions. The results show that, under both LOS and NLOS conditions, the relative gain of CF-mMIMO over Conventional mMIMO first increases and then decreases as the total number of Tx elements grows, reaching its peak when the total number is 64. 

\subsubsection{Spatial Consistency}

Spatial consistency indicates continuous and realistic channel evolution along the UT trajectory in a local area. An early research showed that BER was not only a function of RMS delay spread but was also a function of the temporal and spatial distribution of multipath components as the UT moves. Thus, evaluating system performance should take spatial consistency into account\cite{Ju2018simulating}. 

Based on measurement campaign at 3.75 GHz in industrial scenarios, the work in \cite{nelson2025measurement} presented an accurate, yet simple, spatially consistent channel model, which can be used for realistic system simulations of distributed antenna systems in industrial scenarios. The model is inspired by the COST 2100 framework\cite{liu2012cost},  and the parameters are derived from the measured scenario with distributed single antennas in the environment.

In distributed antenna systems, it is likely that one or more access points (APs) are in LOS, while others are in OLOS. Using Fresnel zone coverage, the obstruction of data acquired in \cite{nelson2025measurement} is classified into two states. One in which obstruction lies between 0\% and 50\%, and another in which obstruction is in the range of 50\% to 100\%. The states are named LOS and OLOS, respectively. 
The matrix $\boldsymbol{H}^{(m)}$ collecting all captured channel transfer functions from the measurement is split into the matrices $\boldsymbol{H}_{\mathrm{LOS}}^{(m)}$ and $\boldsymbol{H}_{\mathrm{OLOS}}^{(m)}$, i.e.,
\begin{equation}
\begin{split}
\boldsymbol{H}^{(m)}=\left[\widetilde{\boldsymbol{h}}_{0}^{(m)}, \ldots, \widetilde{\boldsymbol{h}}_{N_{\mathrm{t}}-1}^{(m)}\right] \in \mathbb{C}^{N_{\mathrm{f}} \times N_{\mathrm{t}}},
\end{split} 
\end{equation}
\begin{equation}
\begin{split}
\boldsymbol{H}_{\mathrm{LOS}}^{(m)}=\left[\widetilde{\boldsymbol{h}}_{k}^{(m)} \mid k \in\left\{0, \ldots, N_{\mathrm{t}}-1\right\}:[\mathbf{L O S}]_{m, k} \leq 50\right],
\end{split} 
\end{equation}
\begin{equation}
\begin{split}
\boldsymbol{H}_{\mathrm{OLOS}}^{(m)}=\left[\widetilde{\boldsymbol{h}}_{k}^{(m)} \mid k \in\left\{0, \ldots, N_{\mathrm{t}}-1\right\}:[\mathbf{L O S}]_{m, k} \geq 50\right] .
\end{split} 
\end{equation}
where $m$, $N_{\mathrm{t}}$ and $N_{\mathrm{f}}$ denotes the number of APs, measured snapshots and tones across the measurement bandwidth, respectively.

After completion of the classification, the spatially consistent channel model is performed following the modeling procedure described in \cite{nelson2025measurement}. While spatially consistent channel modeling has been extensively studied, relatively little attention has been paid to CF-MIMO in new mid-band. Moreover, lack of measurements poses a challenge to accurate spatially consistent channel modeling, especially for CF-MIMO. These remain important topics for future research. 

\subsection{Intelligent XL-MIMO}
\par Intelligent XL-MIMO antenna can automatically adjust the beam direction, shape, and gain according to the real-time changes in the communication environment, thus optimizing the signal transmission quality. Compared with traditional fixed beam antennas, they are more adaptive, directional and have stronger anti-interference capabilities.

\subsubsection{Sparse Antenna Array}

The sparse antenna array is a non-uniform large-scale array structure extended from the framework of sparse sampling technology in the time domain. It can break through the limitations of Nyquist sampling theorem and has significant advantages over uniform array structures in terms of degrees of freedom, mutual coupling of array elements and redundancy. Under the same number of physical array elements, sparse arrays have a larger array aperture and higher degrees of freedom. Sparse arrays can effectively suppress the mutual coupling effect between array elements and the correlation between the noise received by each array element by increasing the spacing between array elements. Under the same physical array aperture, sparse arrays have fewer physical array elements, which helps to reduce array redundancy and the amount of data to be processed.

\par Then, the cross-correlation coefficient between the $m$-th antenna array element and the $n$-th antenna array element is
\begin{equation}
\begin{split}
\rho[m, n] =\int_\phi \int_\theta v_m(\phi,\theta)v_n^*(\phi,\theta)PAS(\phi,\theta)cos(\phi)d\phi d\theta ,
\end{split}
\end{equation}
where $\ast$ is the complex conjugate, and $PAS(\phi,\theta)$ is power angle spectrum of AOD.

\par
\begin{figure}[!htbp]
	\setlength{\abovecaptionskip}{0.1 cm}
	\centering
	\includegraphics[width=0.48\textwidth]{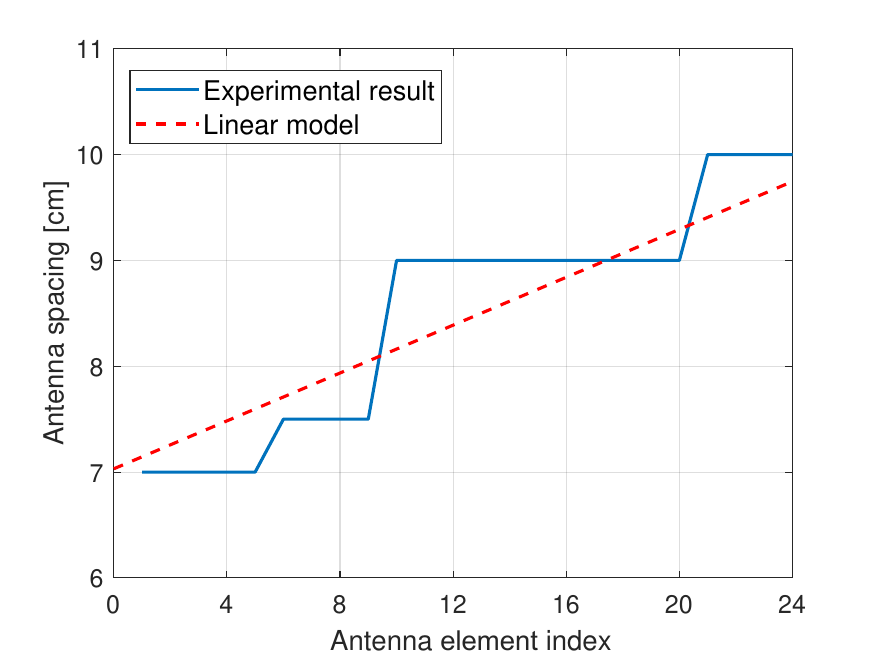}
	\caption{ The antenna spacing of MIMO array elements.}
	\label{sparse_antenna}
\end{figure}

 Antenna movement can provide new degrees of freedom (DoFs) to fully exploit the wireless channel spatial variation. In Fig. \ref{sparse_antenna}, the array spacing of MIMO array is given according to spatial angular spread and correlation coefficient \cite{miao2025far}.

\par According to the spatial non-stationarity of angular spread on the array, the array element spacing can be designed to be non-uniform, that is, the sparse array is used. The array element spacing is arbitrary, but due to the influence of the physical size of the array and mutual coupling effect, it is generally required that the array element spacing is not less than half wavelength, which is satisfied under the correlation parameter settings (the correlation is 0.3) in Fig. \ref{sparse_antenna}. These results are useful for the design and research of movable antenna arrays. For example, a linear model can be used to flexibly optimize the design of an antenna array element spacing with spatial non-stationary characteristic. Besides, depending on the value setting of the spatial correlation, the array spacing can be flexibly designed to effectively take advantage of the spatial non-stationary characteristics. For instance, the design of distributed structure is suitable for Wi-Fi scenarios and mid-band wireless communication.

\subsubsection{Movable Antenna Array}

The movable antenna array has been proposed as a new paradigm to overcome the inherent limitations of fixed-position antenna systems and to better exploit spatial degrees of freedom \cite{Zhu2024MovableAntenna}. By connecting each antenna to a radio frequency (RF) chain via a flexible cable, movable XL-MIMO can be dynamically repositioned within a local spatial region. With multiple movable antennas deployed at the transmitter and/or receiver, spatial diversity can be effectively enhanced through adaptive antenna positioning.

Preliminary studies have demonstrated the potential of movable antennas (MAs) for wireless channel modeling. A representative example is that the work in \cite{shao20246d} proposed a six-dimensional movable-antenna (6DMA) base-station (BS) channel model that explicitly incorporates the 3D position and 3D rotation of each surface. According to the model, the channel between any location of user and the 6DMA-BS can be expressed as
\begin{equation}
\begin{aligned}
\mathbf{h}(\mathbf{q}, \mathbf{u})
&= \sqrt{\nu} e^{-j \frac{2 \pi d}{\lambda}}
\Big[
\sqrt{g\left(\mathbf{u}_{1}\right)}
\mathbf{a}\left(\mathbf{q}_{1}, \mathbf{u}_{1}\right)^{T},
\cdots, \\
&\qquad
\sqrt{g\left(\mathbf{u}_{B}\right)}
\mathbf{a}\left(\mathbf{q}_{B}, \mathbf{u}_{B}\right)^{T}
\Big]^{T}
\in \mathbb{C}^{N B \times 1}
\end{aligned}
\end{equation}
where $\nu$ is the path gain, $d$ denotes the distance between the user’s location and the reference position of the 6DMA-BS, $\lambda$ denotes the carrier wavelength, $N$ denotes the total number of antennas on a 6DMA surface, $B$ denotes the total number of the 6DMA surfaces. $g\left(\mathbf{u}_{1}\right)$  and $\mathbf{a}\left(\mathbf{q}_{1}, \mathbf{u}_{1}\right)$ is the effective antenna gain and the steering vector of the 1st 6DMA surface, respectively\cite{shao20246d}. In particular, $\mathbf{q}$ and $\mathbf{u}$ are used to characterize the 3D position and 3D rotation of the 6DMA surfaces.

\section{Algorithm Evaluation}
\label{sec:V}

The channel estimation, beamforming scheme design, and deep learning-empowered processing are reviewed in this section. Again, due to the extremely large array aperture, signal processing schemes for XL-MIMO systems would involve very high computational complexity. Thus, to promote practical implementation and meet the green communication demands for future communications, low-complexity signal processing schemes for XL-MIMO should be developed. 

\subsection{Conventional Algorithm}

\subsubsection{Channel Estimation}

\begin{itemize}
    \item Near-field channel estimation algorithms
\end{itemize}

In the XL-MIMO system, there exist the non-negligible near-field spherical-wavefront property, which makes the existing low-pilot overhead channel estimation algorithms relying on the channel sparsity in the angular domain ineffective. Therefore, to study the near-field channel characteristics, many studies have been conducted on channel estimation algorithms\cite{Wei2022channelEstimation,Wang2025enhanced,Ghermezcheshmeh2023parametric,Han2020channelEstimation,Cheng2025tensor}.

To capture near-field spherical wave characteristics, the authors in \cite{Wei2022channelEstimation}  represented XL-MIMO channels in the polar domain, integrating angular and distance information to induce channel sparsity. They introduced the on-grid polar-domain simultaneous orthogonal matching pursuit (P-SOMP) algorithm for efficient estimation, alongside an off-grid variant (P-SIGW) for enhanced accuracy. Finally, simulations verified that these approaches outperform traditional angular-domain SOMP and SIGW algorithms.

Building on polar-domain techniques, \cite{Wang2025enhanced} developed an enhanced channel estimation algorithm incorporating lazy residual updates and adaptive weighting. This approach effectively balances computational efficiency with estimation accuracy. Notably, it exhibited superior NMSE performance compared to existing methods, such as angular domain OMP, polar domain OMP, beam split pattern detection (BSPD), bilinear pattern detection (BPD) and multi-candidate BPD (MBPD), under low-SNR conditions, while demonstrating robustness in various system settings.

Focusing on LOS paths in millimeter and terahertz bands, \cite{Ghermezcheshmeh2023parametric} designed a channel estimation scheme grounded in a parametric physical channel model. By leveraging the continuous surface radiation beam structure, the near-field channel was approximated as a superposition of simplified far-field channels. Consequently, the resulting iterative algorithm operates with training overhead and computational complexity independent of the antenna count, demonstrating significant performance gains in scattering-poor environments.

To address near-field spatial non-stationarity, \cite{Han2020channelEstimation} mapped subarrays to scatterers to construct an array response vector representation. The study proposed two distinct estimation methods. One method is a low-complexity subarray-wise approach based on a refined OMP algorithm. Another method is the scatterer-wise approach, where multi-subarray gains are utilized to validate the visible regions of scatterers and accurately pinpoint their locations.

Offering an alternative mathematical perspective, for near-field millimeter-wave XL-MIMO systems, \cite{Cheng2025tensor} introduced a tensor-based channel estimation framework. By analyzing the received space-time-frequency (STF) tensor, the authors derived an alternating least squares (ALS) algorithm for CANDECOMP / PARAFAC (CP) decomposition. Theoretical analysis and simulations jointly validated the applicability of this tensor framework in short-propagation MIMO scenarios, highlighting its capacity to significantly improve estimation accuracy while minimizing pilot overhead.

\begin{itemize}
    \item Hybrid-field Channel Estimation Algorithms
\end{itemize}

As mentioned earlier, traditional far-field channel estimation can be carried out through low-overhead estimation based on compressive sensing that relies on angular domain sparsity. However, near-field channel estimation requires simultaneous consideration of angular and distance information in the polar domain, making the near-field channel also exhibit sparsity, thereby fully capturing the near-field spherical baud property. However, in actual scenarios, scatterers may exist simultaneously in both the far-field and near-field regions, forming a hybrid-field propagation environment. How to conduct channel estimation for the hybrid field is currently a major challenge in the research of XL-MIMO systems. Some studies have been conducted on hybrid-field channel estimation \cite{Wei2022channelEstimation,Jiang2025joint}.

To address the coexistence of far-field and near-field path components, \cite{Wei2022channelEstimation} developed a hybrid-field OMP (HF-OMP) channel estimation scheme within the polar domain. By concurrently leveraging hybrid-field channel characteristics and polar-domain sparsity, this approach effectively extracts both path components. The simulation results showed that the HF-OMP algorithm achieves superior NMSE performance compared with the far-field and near-field OMP algorithms.

Shifting towards a data-driven perspective, \cite{Jiang2025joint} introduced a joint dictionary learning and channel estimation framework based on the alternating direction method of multipliers (ADMM). This method extracts essential hybrid-field characteristics directly from extensive channel measurement data to construct an adaptable dictionary and achieved accurate channel estimation. Consequently, it consistently outperforms existing schemes constrained by fixed far-field or near-field dictionaries.

\subsubsection{Beamforming}

\begin{itemize}
    \item Near-field beamforming algorithms
\end{itemize}

The conventional linear beamforming practical schemes for mMIMO systems mainly fall into three categories: maximal ratio (MR), ZF, and minimum mean-square error (MMSE). Due to the extremely large array scale of XL-MIMO systems, the conventional assumption based on far-field uniform plane wave (UPW) is no longer suitable for its physical characteristics. To address this issue, some works have conducted research on near-field modeling and the optimization of corresponding beamforming algorithms \cite{Li2022near-field, Guerra2022clustered, Li2023modular, Lu2022near-field}.

As a viable solution to massive antenna deployment in XL-MIMO, the modular extremely large-scale array is a new antenna array architecture based on near-field non-uniform spherical wave (NUSW) that has attracted considerable research attention. In this context, \cite{Li2022near-field} and \cite{Li2023modular} respectively investigated the mathematical modeling and conducted the performance analysis of beamforming for modular XL-array communications, deriving a closed-form expression for the maximum SNR and the optimal maximum ratio combining (MRC) beamforming by precisely modeling the signal amplitude and phase, as well as projecting apertures across all modular elements. The maximum SNR depends on key system parameters such as the overall modular array size, the distances between adjacent modules along all dimensions, and the positions of users. The SNR based on the UPW far-field model increases linearly and unboundedly with the growth of the number of modules, whereas the SNR of the modular XL-MIMO under the NUSW near-field model gradually converges to a constant upper bound as the number of modules along a certain dimension increases, which highlights the importance of near-field modeling for modular XL-MIMO array communications. 

To capture spatial-time evolution in XL-MIMO systems, \cite{Guerra2022clustered} proposed a double-scattering channel model encompassing both BS and UE scattering clusters. Assuming ULA and UPA topologies at the BS, the authors comprehensively evaluated key performance metrics, namely signal-to-interference-plus-noise ratio (SINR), condition number (CN), and SE, under MRC, ZF, and MMSE schemes. The analysis revealed that MMSE consistently outperforms MRC and ZF across all evaluated large-scale arrays. Particularly in densely crowded configurations, MMSE demonstrated exceptional robustness, theoretically supporting an unbounded user capacity.

Focusing on near-field multi-user communications, \cite{Lu2022near-field} employed a projected aperture non-uniform spherical wave (PNUSW) model to analyze the SNR across typical MRC, ZF, and MMSE beamforming schemes. Crucially, their investigation uncovered a new DoF for inter-user interference (IUI) suppression, which is achieved through distance separation along the same angular direction. Extensive simulations corroborated these findings, demonstrating the effectiveness of utilizing the distance dimension to suppress interference in multi-user XL-MIMO systems.

\begin{itemize}
    \item Far-field beamforming algorithms
\end{itemize}

To address the topological challenges posed by the massive proliferation of antennas in conventional mMIMO, \cite{Bjornson2019utility} investigated far-field transmissions over a large intelligent surface (LIS). By comparing different precoding schemes, they demonstrated that ZF precoding outperforms MR transmission for practically sized surfaces, although this performance gap diminishes as the LIS dimensions expand. Furthermore, the study showed that ZF enables highly efficient power allocation tailored to different utility functions. Notably, they established that although an ideal LIS is a continuous surface, its beam pattern is closely approximated when using discrete antennas of size $\lambda/4 \times \lambda/4$.

\begin{itemize}
    \item Hybrid-field beamforming algorithms
\end{itemize}

To establish a rigorous mathematical framework for general XL-array/surface wireless systems, \cite{Lu2022communicating} introduced a unified modeling approach applicable to both discrete arrays and continuous surfaces. This model accounts for spatial variations in signal phase, power, and projected aperture across array elements. With the optimal MRC/maximum ratio transmission (MRT) beamforming, a closed-form SNR expression was derived for single-user up-link/down-link communication with 3D user directions. This expression indicates that, instead of scaling linearly with the antenna number M as in conventional UPW modeling, the SNR with the more generic model increases with M with diminishing return, which is governed by the collective properties of the array, such as the array occupation ratio and the physical sizes of the array along each dimension, while irrespective of the properties of the individual array element.

\subsection{Algorithms Based on Large Models/AI}

\subsubsection{Channel Estimation}

\begin{itemize}
    \item Near-field channel estimation algorithms
\end{itemize}

Capitalizing on the inherent polar-domain sparsity of the near-field channel, \cite{Lei2024channelEstimation} proposed a polar-domain multiple residual dense network (P-MRDN) for XL-MIMO systems. To further elevate estimation accuracy, the authors designed a polar-domain multi-scale residual dense network (P-MSRDN). By synergizing polar-domain sparsity, deep residual learning, and multi-scale feature extraction, these network designs achieve superior generalization capabilities. Simulation results showed that the NMSE performance of the proposed P-MRDN and P-MSRDN is superior to the MRDN scheme and the conventional CS scheme.

\begin{itemize}
    \item Hybrid-field channel estimation algorithms
\end{itemize}

To streamline hybrid-field THz UM-MIMO channel estimation, \cite{Yu2023anAdaptive} introduced a low-complexity general Deep learning (DL) framework. This framework utilized existing iterative channel estimators and employed fixed-point iteration to calculate the channel estimation. Each iteration is implemented by a fixed point network (FPN). Additionally, a specific FPN-enhanced algorithm based on orthogonal approximate message passing (OAMP), named FPN OAMP, was proposed to model neural networks with any depth and adapt to the hybrid-field channel conditions. Simulation results demonstrated that the proposed method showed significant gains in various key performance indicators. The computational complexity of the proposed FPN-OAMP algorithm is adaptive and has strong robustness to distribution changes, making it more suitable for future wireless networks with complicated hybrid-field channel conditions.

\subsubsection{Beamforming}
\begin{itemize}
    \item Near-field beamforming
\end{itemize}

In scenarios with an extremely large number of antennas, to meet the requirements of high throughput and ultra-low latency, efficient, high-performance, and low-complexity detection algorithms for massive/massive MIMO are needed. Some works investigated low-complexity algorithms that utilize AI for data processing \cite{He2023GNN,Liu2024double-layer}.

To enhance detection capabilities with minimal computational overhead, \cite{He2023GNN} introduced AMP-GNN, a model-driven deep learning detector based on a graph neural network (GNN)-enhanced AMP algorithm. The neural network structure was obtained by unfolding the AMP detector and incorporating the GNN module.The simulation results showed that the AMP-GNN algorithm achieved the low complexity of the AMP detector and the efficiency of the GNN module, significantly improving the performance of the AMP detector.

Focusing on near-field communications, \cite{Liu2024double-layer} analyzed the uplink SE of CF XL-MIMO systems, wherein both base stations and user equipment are configured with XL-MIMO panels. To address the ensuing high-dimensional signal processing challenges, the authors designed a multi-agent reinforcement learning (MARL)-based power control algorithm that incorporates predictive management and distributed optimization architecture. The results showed that the proposed MARL-based algorithm effectively strikes a balance between spectral SE performance and convergence time.

\subsection{Algorithms with Application Potential}

\subsubsection{Channel Estimation}

\begin{itemize}
    \item Near-field channel estimation algorithms
\end{itemize}

Starting from the actual deployment requirements in the industry, the balance between low complexity, low overhead and scheme performance is an important reference standard for the implementation of XL-MIMO channel estimation schemes. Some works studied channel estimation schemes based on low complexity \cite{Demir2022channelModeling,Zhu2021bayesian}, which showed excellent performance compared with traditional schemes. 

To capture the effects of non-isotropic scattering and directive antennas, \cite{Demir2022channelModeling} formulated a holographic XL-MIMO channel model based on UPAs. Building upon this model, the authors introduced a reduced-subspace least squares (RS-LS) channel estimation scheme. This approach exploits the partial correlation induced by the array geometry to identify a reduced-rank subspace that encompasses the feature space of any arbitrary spatial correlation matrix. The simulation results showed that the performance of this scheme is superior to that of the traditional LS estimator. The core advantage lies in that it does not rely on the channel statistics prior information of specific users, avoiding the overhead in aspects such as the acquisition and update of statistical information caused by the mobility of UE. In addition, its low complexity feature is suitable for large-scale antenna deployment scenarios.  

In the context of multi-user multiple-input single-output (MU-MISO) OFDM broadband systems equipped with extremely large antenna arrays (ELAAs), \cite{Zhu2021bayesian} developed a structured prior based on a Hidden Markov Model (HMM) to improve the structured sparsity of spatially non-stationary ELAA channels. Leveraging this framework, the authors designed a turbo OAMP algorithm to facilitate low-complexity Bayesian inference. Compared to state-of-the-art alternatives, this algorithm not only achieves superior NMSE performance under low SNR conditions, but also drastically reduces the required pilot overhead.

\begin{itemize}
    \item Hybrid-field channel estimation algorithms
\end{itemize}

To minimize pilot overhead in partially-connected hybrid beamforming architectures for UPA, a tailored hybrid-field channel estimation scheme was developed \cite{Ruan2024low}. By optimizing the hybrid beamforming architecture, namely the phase shift matrix optimization strategy, precise angular domain recovery was achieved. Additionally, the proposed spatial filtering technique then achieves effective field separation: far-field components are extracted through joint analysis of beamwidth characteristics and amplitude distributions, while residual near-field components are subsequently recovered via compressed sensing-based reconstruction. The simulation results showed that this algorithm can reconstruct the hybrid-field channel estimation with a 50\% less pilot overhead.

\subsubsection{Beamforming}

\begin{itemize}
    \item Near-field beamforming
\end{itemize}

Aiming to bridge the gap between computational efficiency and optimal performance, three accelerated randomized Kaczmarz (RK) methods were introduced based RZF receiver designs: RK-RZF, greedy RK (GRK)-RZF, and randomized sampling Kaczmarz (RSK)-RZF \cite{Croisfelt2021accelerated}. Among these variants, the RK-RZF algorithm strikes the most favorable performance-complexity trade-off, yielding complexity reductions of 20\% in typical mMIMO setups and a substantial 70\% in XL-MIMO systems. Alternatively, the GRK-RZF scheme exploits the full residual information of the systems of linear equation (SLE) to accelerate convergence, making it more suitable for extreme cases where IUI and sparsity effects cannot be ignored. Conversely, while the RSK-RZF approach operates with significantly less residual information, it incurs a noticeable performance penalty.

Addressing the novel communication paradigms of massive MIMO systems equipped with extremely large-scale antenna arrays, \cite{Amiri2018extremely} investigated various low-complexity data detection algorithms. The authors proposed a linear data fusion method, as well as a graph-based algorithm inspired from coded random access which uses low complexity and distributed scheme for data detection. This method converts the channel propagation environment into a bipartite graph and detects the users in a novel scheme.

\subsection{Performance Evaluation in the New Mid-Band}
Regarding the spatial geometry configuration, it is assumed that a single-user terminal is located at a distance of $r = 20$ m from the array center, with the elevation and azimuth angles of signal arrival set at $\theta = 60^\circ$ and $\phi = 45^\circ$, respectively. In terms of the physical architecture of the array, the system is equipped with a total of 768 antenna elements arranged in $16 \times 16$ modular layout (with each module containing $M=3$ antennas), where the antenna element spacing is tied to the half-wavelength.

\subsubsection{Different Frequencies}
We investigated the SNR of the UPW and NUSW models with 768 antennas (16×16×3) in the sub-6 GHz, and 6-24 GHz band.
\begin{figure}[!htbp]
        \vspace{0mm}
	\setlength{\abovecaptionskip}{0.1 cm}
	\centering
	\includegraphics[width=0.45\textwidth]{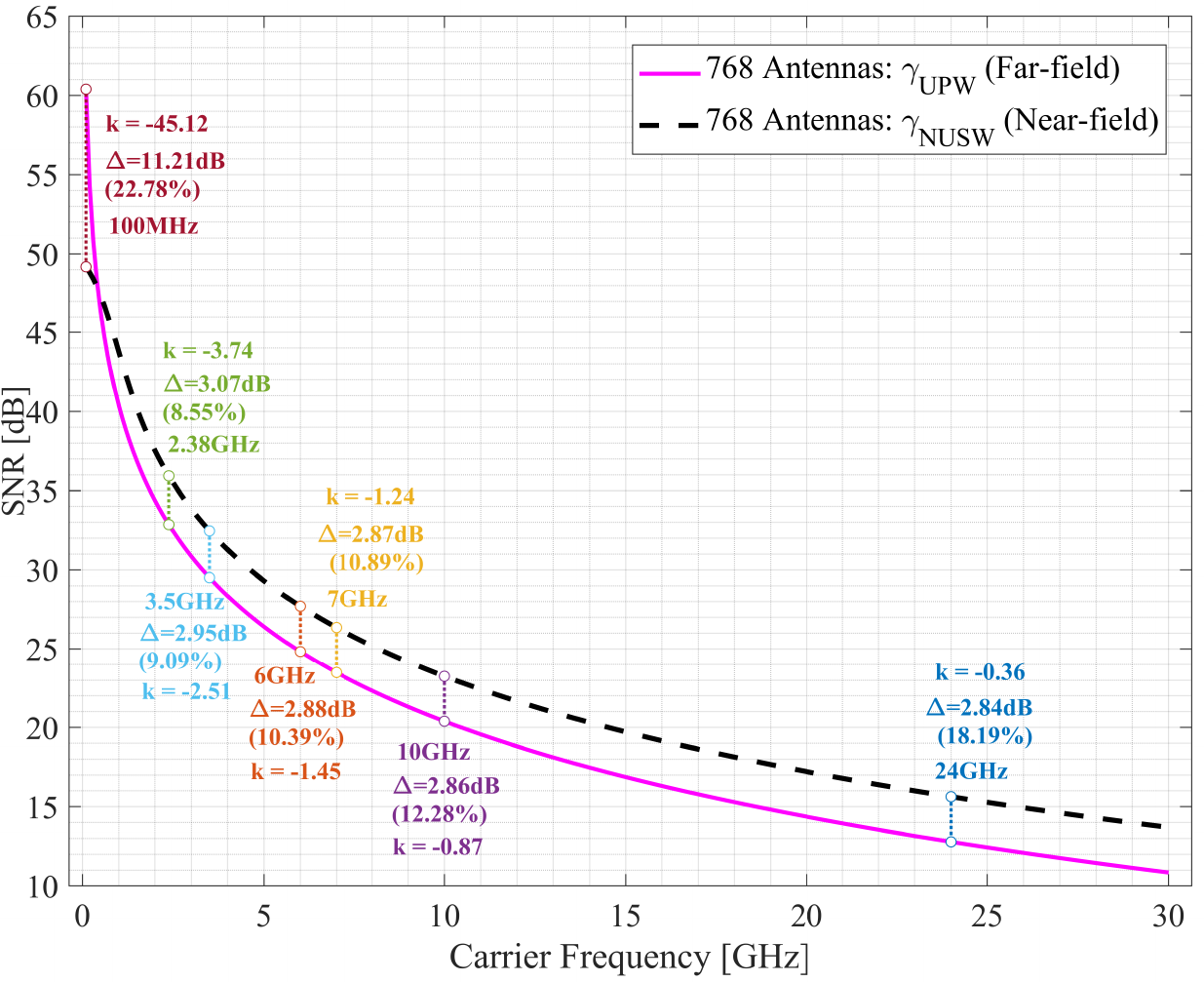}
	\caption{Comparison of SNR with 768 array elements from sub-6 GHz to mmWave band.}
	\label{frequency1-2}
\end{figure}
As illustrated in Fig. \ref{frequency1-2}, the SNR exhibits an overall monotonic decline as the frequency increases. Whether employing the far-field UPW model or the near-field NUSW model, the system SNR consistently degrades with rising frequency. This occurs because, under a half-wavelength spacing configuration, the effective receiving aperture of a single antenna is proportional to the square of the wavelength ($A_e \propto \lambda^2$). Higher frequencies yield shorter wavelengths, thereby shrinking the effective receiving aperture of individual antennas. Consequently, the total effective area of the array available for capturing spatial energy is reduced.

As the frequency increases, the attenuation slope ($k$) of the system gradually flattens. In the low-frequency band (e.g., 2.38-3.5 GHz), even a marginal increment in frequency causes a severe degradation in SNR, manifesting as a highly steep slope. However, upon entering the mid-frequency or millimeter-wave bands, the attenuation becomes increasingly gradual in the logarithmic domain, with $k$ approaching 0. From a physical perspective, the effective aperture of the receiving antenna scales as $A_e \propto \lambda^2$ (i.e., inversely proportional to the square of the frequency). As the frequency progressively rises, $A_e$ drops rapidly at first before declining more slowly; correspondingly, the total effective area of the array used to capture spatial energy undergoes a rapid initial reduction followed by a slower decrease. From a mathematical standpoint, free-space path loss is proportional to the square of the frequency. In the logarithmic domain, $SNR \propto  -20\log_{10}(f)$, and the derivative of this logarithmic function with the slope $k\propto -1/f$. Therefore, for a given absolute frequency increment, a higher base frequency results in a smaller relative ratio of change, which translates to a reduced additional dB loss in the logarithmic domain.

Specifically, at the 100 MHz frequency point, the traditional far-field UPW model completely breaks down. This is because, at this specific frequency, the half-wavelength antenna spacing increases drastically, causing the physical aperture of the 768-element array to expand to a scale of hundreds of meters. At this point, the user-to-array distance is simply set to 20 m, yet the far-field UPW formula still relies on the ideal assumption that the propagation distances from all antennas to the user are approximately equal (20 m). This assumption completely ignores the severe path loss and phase shifts experienced by the signals traveling from the edge antennas of the massive array to the user, leading to a severe overestimation of the SNR at this frequency.

\subsubsection{Different Numbers of Antennas}
We investigated the variation of SNR with frequency under the near-field NUSW scenario for two different antenna array sizes, including 1536 (16×16×6) and 768 (16×16×3). 

\begin{figure}[!htbp]
	      \vspace{0mm}
      \setlength{\abovecaptionskip}{0.1 cm}
	\centering
	\includegraphics[width=0.45\textwidth]{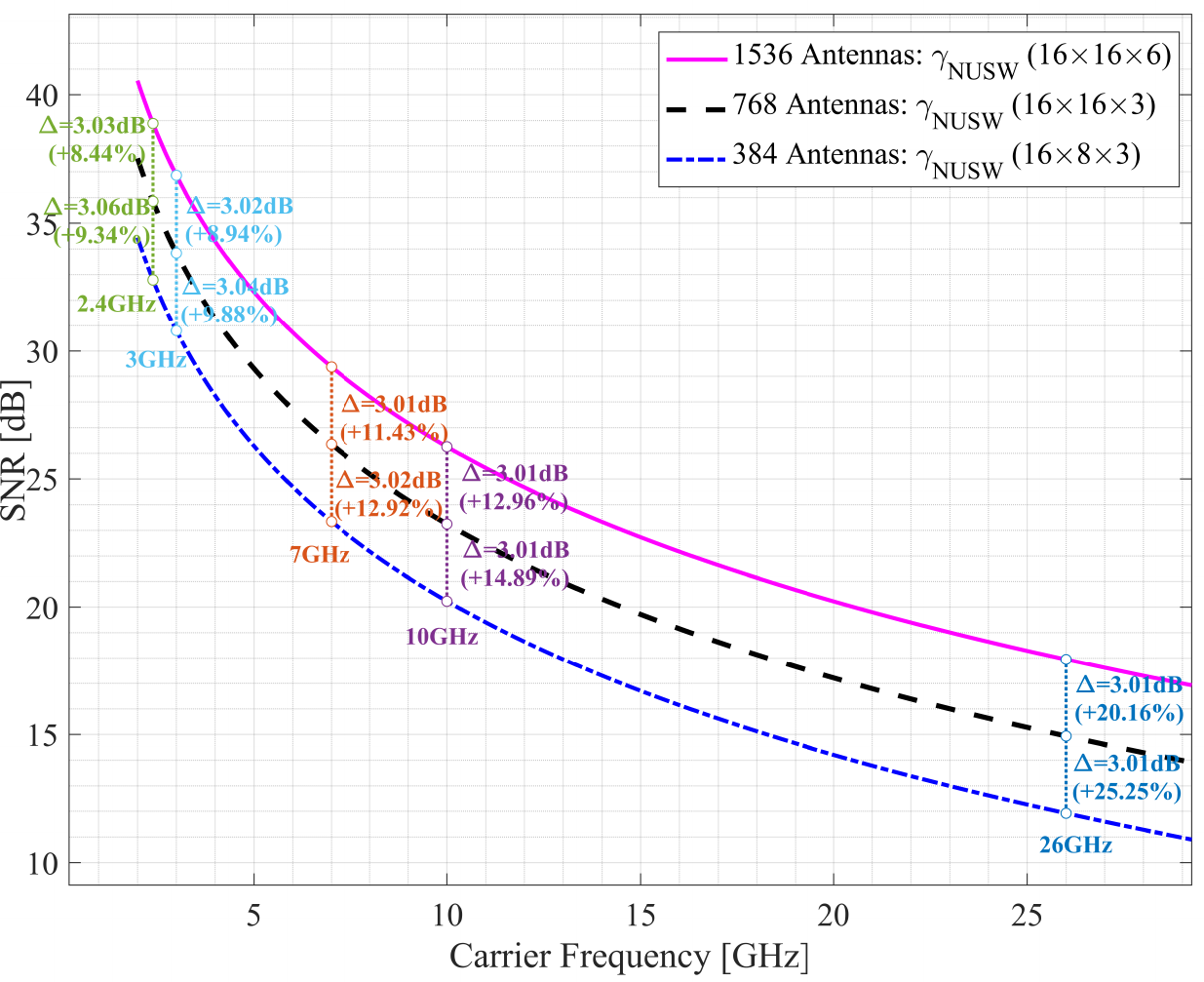}
	\caption{Comparison of SNR with 384, 768, and 1536 array elements from sub-6 GHz to mmWave band.}
	\label{antenna_number2}
\end{figure}

As illustrated in Fig. \ref{antenna_number2}, the achievable SNR for all three configurations exhibits a declining trend with increasing frequency. Notably, the SNR curves of the three antenna schemes remain largely parallel across the entire 2 GHz to 30 GHz frequency band. At the sampled frequency points, the SNR gap between adjacent configurations stabilizes at approximately $\Delta = 3.01$ dB. This observation is highly consistent with the theoretical array gain of roughly 3 dB achieved by doubling the number of antennas.

\begin{figure}[!htbp]
	\setlength{\abovecaptionskip}{0.1 cm}
	\centering
	\includegraphics[width=0.48\textwidth]{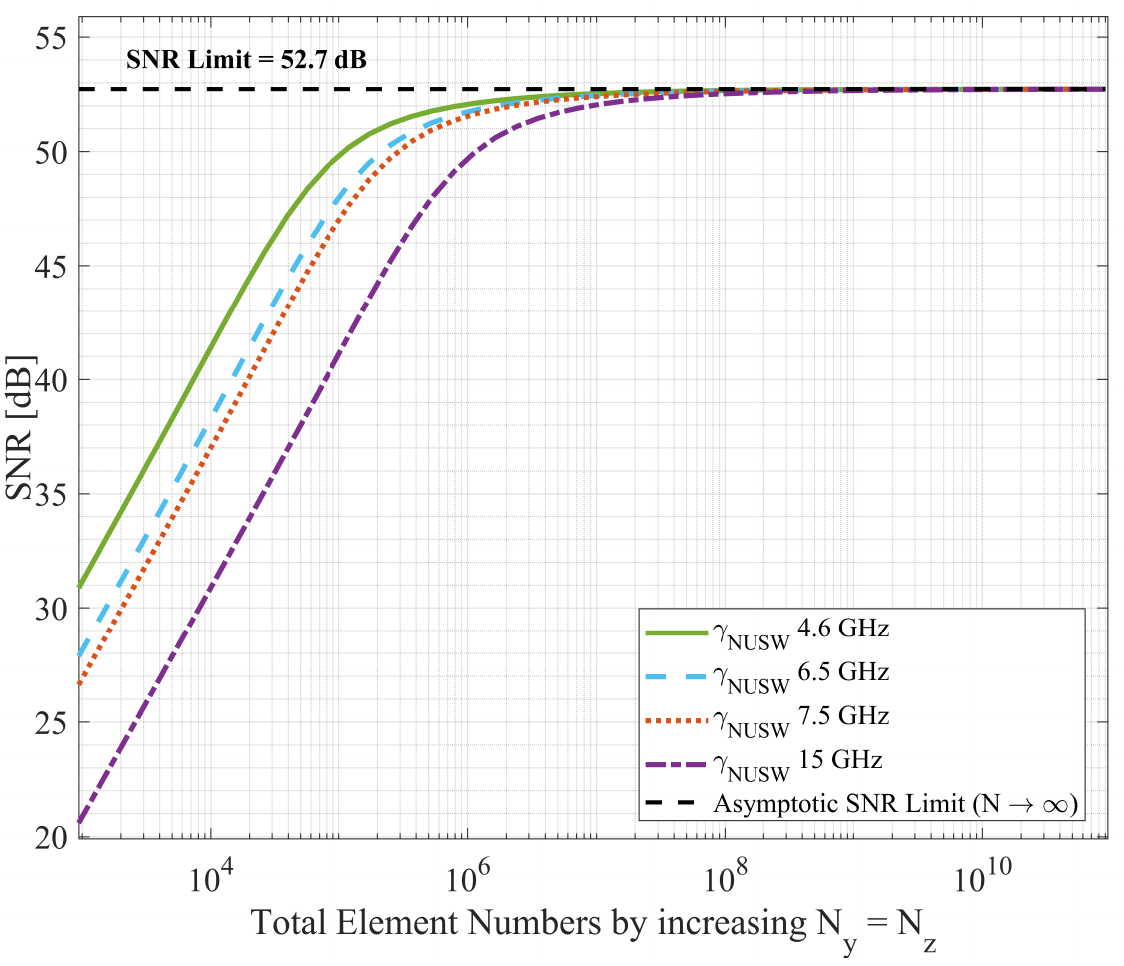}
	\caption{Comparison of SNR under different frequency bands as the number of antennas changes.}
	\label{SNR_antenna_num}
\end{figure}

As illustrated in Fig. \ref{SNR_antenna_num}, the SNR curve generally increases with the number of antennas and ultimately converges to a constant value. When the number of antennas is small, the SNR exhibits rapid growth as the number of antennas increases. This is because the electromagnetic waves impinging on the array can be approximated as far-field UPW. Under the plane-wave assumption, the signal power received by all antenna elements is nearly identical, differing only in phase. Consequently, as the number of antennas increases, the system achieves significant spatial array gain, and the collected signal energy is approximately proportional to the total number of antennas. Conversely, when the number of antennas is large, the growth rate of the SNR decelerates. This is attributed to NUSW propagation. Under the spherical wave model, on one hand, an increase in the number of antennas gradually enlarges the distance between the user and the antennas located at the edges of the array. This exacerbates the free-space path loss, causing the signal energy received by these edge antennas to progressively attenuate. On the other hand, the angle of incidence at the edge antennas, relative to the arrival direction of the electromagnetic waves, also increases. This leads to a reduction in the effective projected area available for the antennas to capture energy. The combination of these two factors causes the growth rate of the SNR to gradually slow down.

For a given number of antennas, the SNR curve for the higher frequency bands within the new mid-band shifts downwards overall. This occurs because, at higher frequency ranges within the new mid-band, the elevated carrier frequency results in a shorter wavelength. Subsequently, this reduces the total physical receiving aperture of the array, decreasing the system capability to capture spatial RF energy, which ultimately leads to an overall degradation in the received SNR.

\section{Field Trials}
\label{sec:VI}
To verify the performance of XL-MIMO antenna array in real communication systems, this section mainly introduces the prototype system incorporating XL-MIMO, including its components, functions, configuration, and field trials.

\subsection{Platform Architecture}
\begin{figure}[!htbp]
	\setlength{\abovecaptionskip}{0.1 cm}
	\centering
	\includegraphics[width=0.46\textwidth]{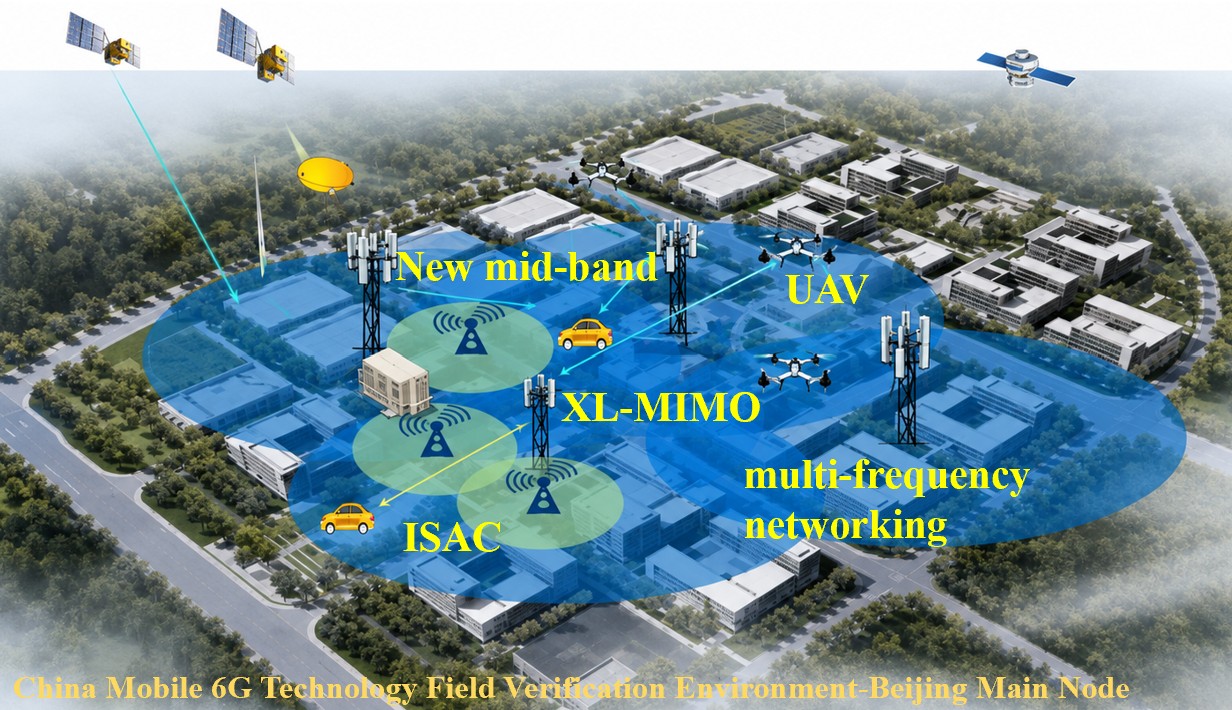}
	\caption{The 6G open innovation test device from prototype verification to large-scale testing.}
	\label{platform_prototype}
\end{figure}

The 6G open innovation test device is composed of three layers: the 6G open network simulation platform, the prototype system, and the field test environment. Among them, the prototype system serves as the key intermediate layer connecting the simulation and the physical implementation, and adopts the ``base + center + core" architecture. The base is a heterogeneous hardware cloud platform (including CPU/GPU/FPGA), enabling on-demand sharing and unified scheduling of computing power; the core modules are modularized to support multiple capabilities such as communication, perception, computing, and artificial intelligence (integrated perception computing intelligence); the core is achieved through service-oriented technology to realize cross-domain collaboration and intelligent orchestration. As shown in Fig. \ref{platform_prototype}, this system is a real end-to-end communication system, supporting concurrent services and seamless integration with real services, and features such as functional plug-in and network customization.

\begin{figure}[!htbp]
	\setlength{\abovecaptionskip}{0.1 cm}
	\centering
	\includegraphics[width=0.46\textwidth]{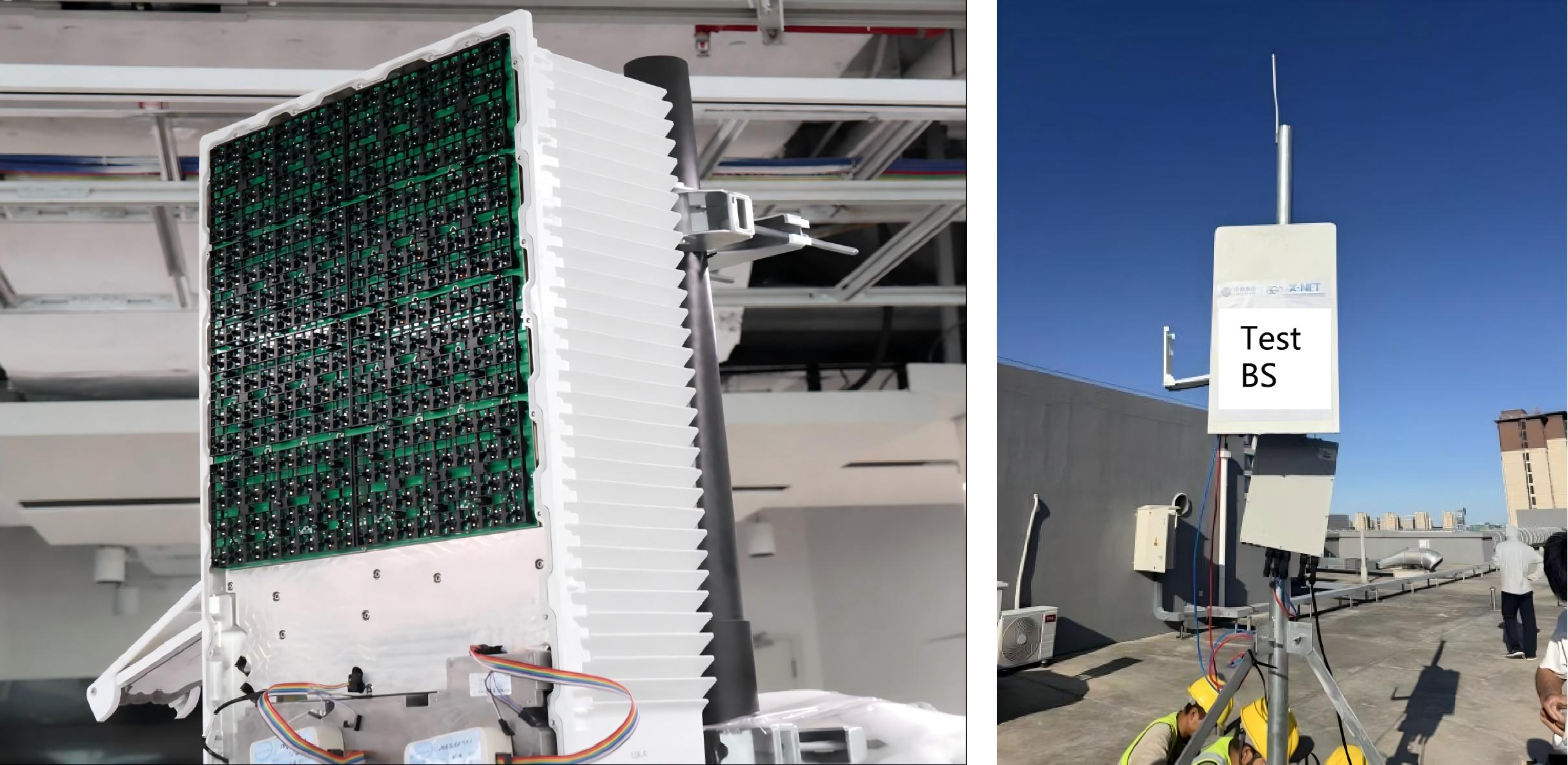}
	\caption{U6GHz band XL-MIMO array antenna base station.}
	\label{XL-MIMO}
\end{figure}

For ISAC scenario, the system can achieve an accuracy of sub-meter level for the perception of mobile targets such as drones and vehicles, with a transmission rate of 6.6 Gbps, which can support intelligent traffic management. In terms of endogenous AI, lightweight model compression reduces the transmission data volume by 80\%, heterogeneous computing power enables microsecond-level dynamic scheduling, meeting the low latency requirements for remote surgeries, etc. At the network self-intelligence level, resource allocation and fault prediction are achieved through feature learning. In new mid-band and millimeter-wave frequency bands, XL-MIMO and beamforming are integrated to support 10-100 Gbps transmission. The XL-MIMO module in the prototype system is equipped with an AAU consisting of 128 channels and 1024 antenna elements in Fig. \ref{XL-MIMO}. Under the conditions of U6GHz, 400 MHz bandwidth, and 8-stream transmission, its peak speed exceeds 9 Gbps. Through the universal backhaul module and the heterogeneous open architecture, the system supports dynamic switching and integrated access of multiple frequency bands such as U6GHz, new mid-band, mmWave, THz, and visible light, and opens up the basic capabilities of the air interface. It provides a comprehensive verification environment for 6G cutting-edge technologies such as integrated perception, intelligent super surface beamforming, and new waveform design, with high and low frequency coordination.

The open test environment is built on the basis of the prototype system and is driven by requirements and applications. It can flexibly set up various network and business scenarios, supporting the verification from technical feasibility to industrial feasibility. Currently, a multi-node connection across Beijing, and other regions is achieved as the main node. In the future, the ``1 + 3 + N" station network across the country is planned to form a large-scale test and verification network. The capability open platform serves as a unified interface, providing five open capabilities, including data, functions, interfaces, environment, and business, and is open for use by all parties involved in research, education, and industry. This device constitutes a complete 6G collaborative innovation platform from prototype verification to large-scale testing and then to capability opening.

\subsection{Prototype System}

The 6G prototype system, as the intermediate layer, serves as a bridge connecting the simulation theory and the physical implementation in the U6GHz (6425-7125 MHz) band. Its most significant feature is the seamless integration of multiple dimensions of capabilities such as communication, perception, computing, artificial intelligence, and security. It also adopts the fronthaul module (supporting ultra-high-speed CPRI interface and flexible variable intermediate frequency signal transmission) to achieve efficient connection with multi-band RF frontends such as new mid-band, millimeter wave, and terahertz.

\begin{figure}[!htbp]
	\setlength{\abovecaptionskip}{0.1 cm}
	\centering
	\includegraphics[width=0.46\textwidth]{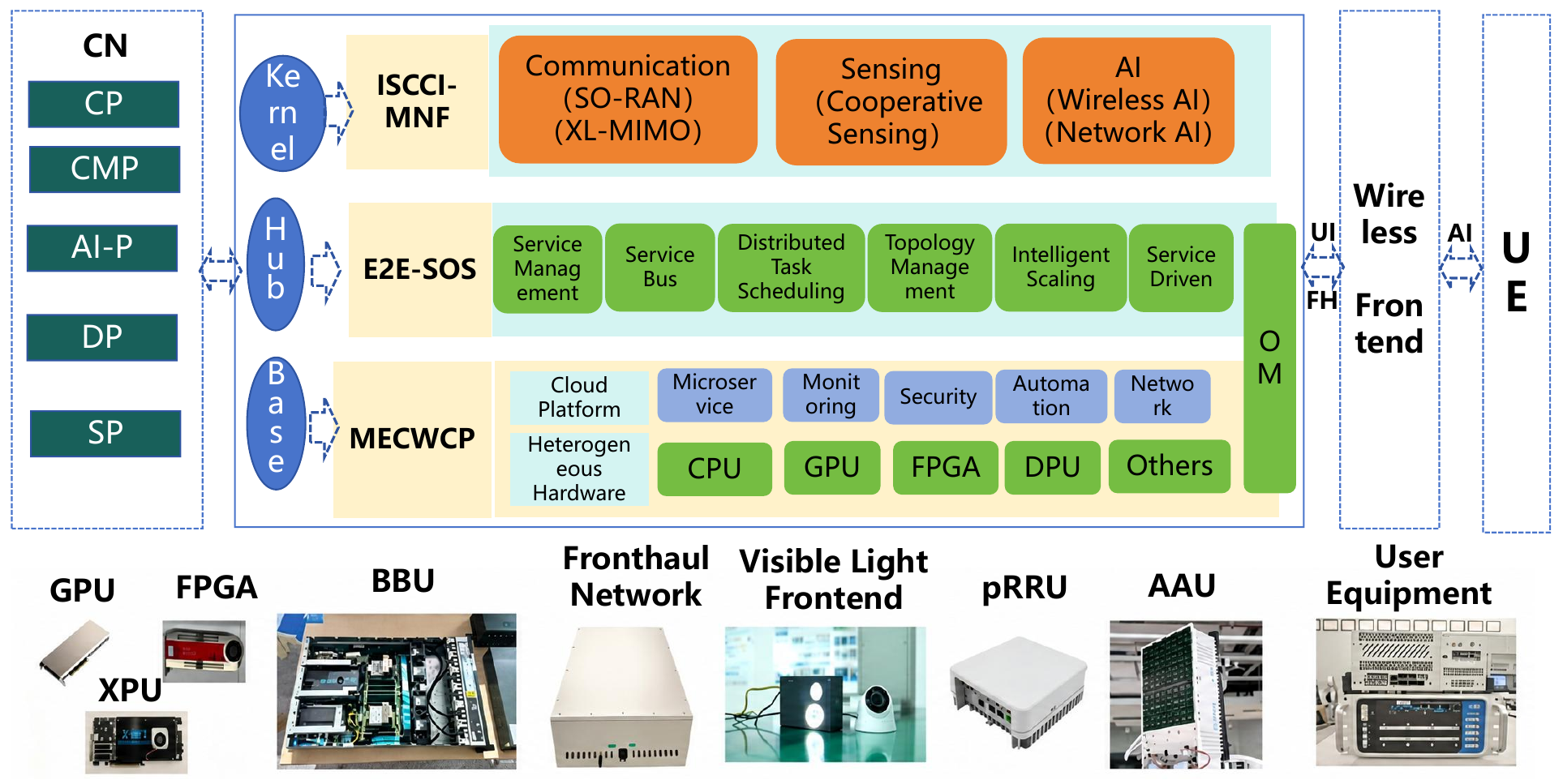}
	\caption{Overall architecture of 6G prototype system platform.}
	\label{architecture}
\end{figure}

An open system architecture based on a wireless cloud platform and heterogeneous hardware, with the core of integrated perception, computing and intelligence functions, and the central role of wireless network serviceization, has created a 6G end-to-end prototype system platform that supports the endogenous integration of multiple elements such as ``communication + sensing + computing + intelligence + X". The overall architecture is shown in Fig. \ref{architecture}.

The protocol stack of the software architecture system adopts the CU and DU separation architecture, and is generally divided into three modules: CU, DU, and PHY. The baseband processing unit (BBU) of the hardware architecture adopts a heterogeneous hardware platform of ``general server + GPU card + accelerator card", using general processors to handle complex logic and upper-layer protocols, and using dedicated accelerators to handle high-throughput and high-real-time tasks of the physical layer.

The XL-MIMO system is equipped with the 1024-element antenna array, and its transceiver chain supports calibration for both transmission and reception. It operates over a carrier frequency range of 6.425-7.125 GHz, with a system bandwidth of 400 MHz, and supports 256-order modulation. In terms of standard support, the system can conduct certain prospective verification activities on the basis of compliance with 3GPP standards.

\begin{table}[htbp]
	\centering
	\caption{Performance of prototype system}
	\label{prototype}
	\renewcommand{\arraystretch}{1.5}
	 \setlength{\tabcolsep}{6 mm}		
	\begin{tabular}{c|c}
	 \hline \hline
	    Parameter    & Performance  \\ \hline
            Work frequency [GHz]    &  6.425-7.125 \\ \hline
            Array element number   &  1024 \\ \hline
            Max bandwidth [MHz]    &  400  \\ \hline
		Beamforming accuracy    &  99.2\% \\ \hline
            Modulation mode   &  256QAM \\ \hline
            Duplex mode   &  TDD \\ \hline
		Single-user peak rate [Gbps]    & 9.1  \\  \hline
            Sensing accuracy &  sub-meter \\  \hline
		Detection range [m]   & 500\\   \hline 
            Angular accuracy [$^\circ$]      &  0.5  \\ \hline \hline
	\end{tabular}
\end{table}

In Table \ref{prototype}, in terms of the openness of the basic capabilities of the wireless air interface, multi-frequency integration access verification capabilities such as U6GHz, new mid-band, mmWave, and THz will be enabled. Through heterogeneous hardware open architectures, multi-standard and multi-interface data transmission will be achieved. Key verification will focus on XL-MIMO antenna array beamforming capabilities and integrated communication-sensing technologies, while also conducting experiments on new waveform design.

\subsection{Performance Evaluation}

\begin{figure}[!htbp]
	\xdef\xfigwd{\columnwidth}
	\setlength{\abovecaptionskip}{0.1 cm}
	\centering
	\begin{tabular}{cc}
		\includegraphics[width=4.3cm,height=4.0cm]{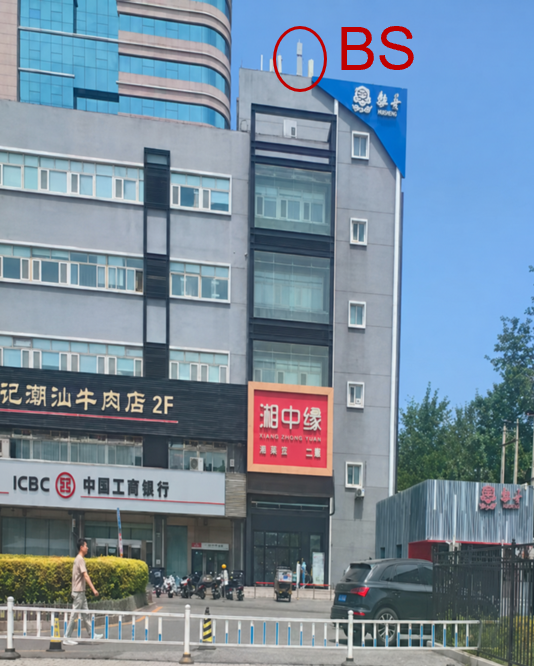} \hspace{-4mm}
		& \includegraphics[width=4.3cm,height=4.0cm]{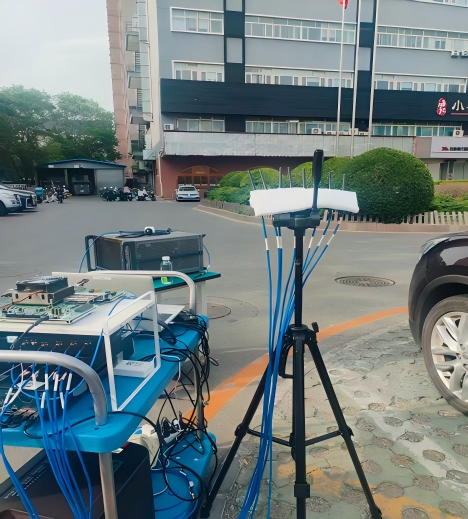}\\
		{\footnotesize\sf (a)} &	{\footnotesize\sf (b)} \\	
	\end{tabular}
	\caption{U6GHz XL-MIMO test site environment. (a) Tx end with BS. (b) Rx end.}
	\label{site_environment}
\end{figure}

As shown in Fig. \ref{site_environment}, field trials of XL-MIMO performance are conducted in an outdoor UMa scenario. The selected frequency band was 6425-6825 MHz, with a bandwidth of 400 MHz. The XL-MIMO base-station antenna array was deployed on a rooftop at a height of approximately 22 m, while the receiver was located at ground level and consisted of an 8-element signal receiving array.

\begin{figure}[!htbp]
	\setlength{\abovecaptionskip}{0.1 cm}
	\centering
	\includegraphics[width=0.48\textwidth]{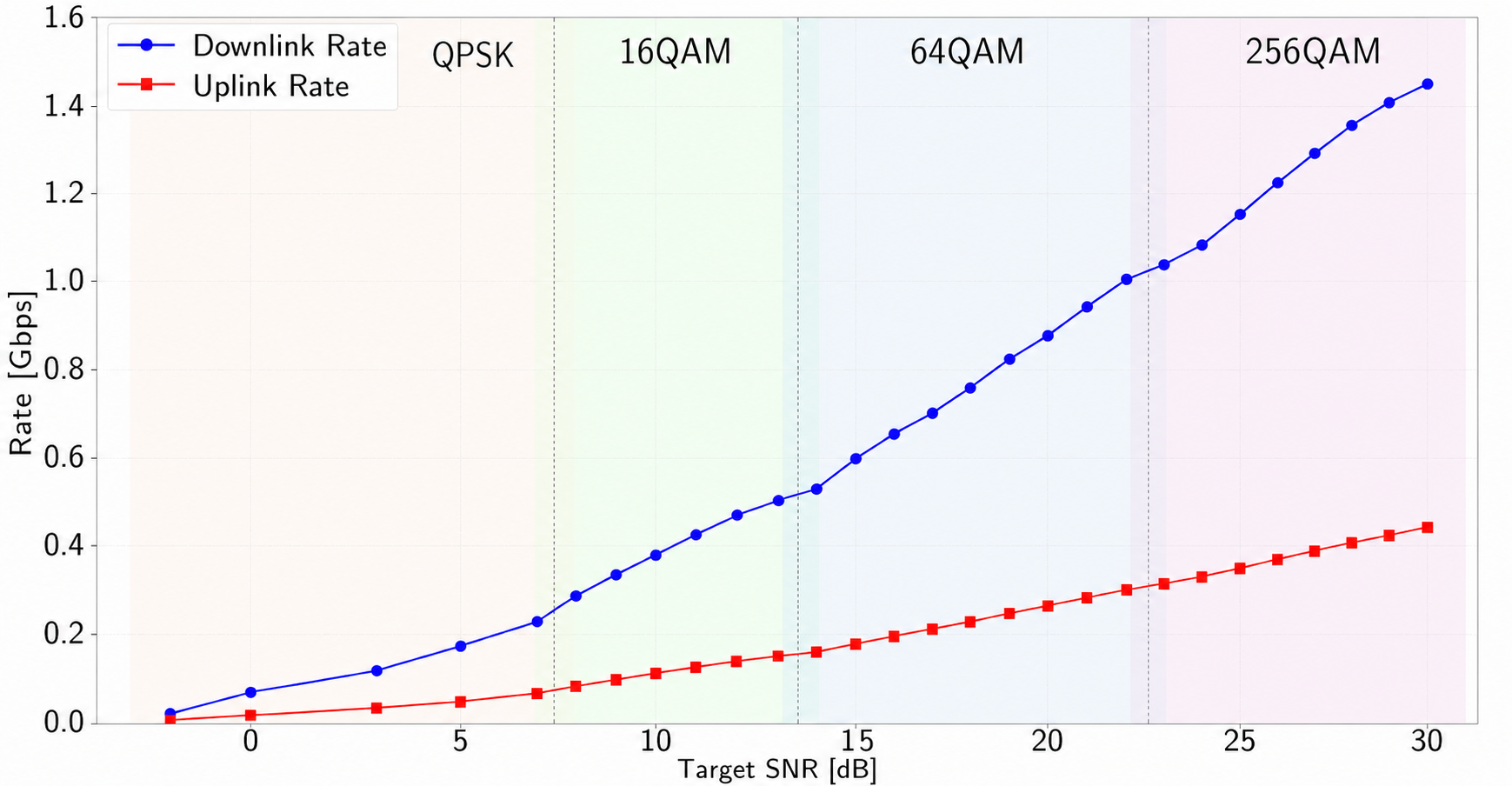}
	\caption{The single-stream rate and SNR of XL-MIMO in the U6GHz band with 400 MHz bandwidth (Modulation mode switching regions indicated).}
	\label{test_result}
\end{figure}
The Fig. \ref{test_result} shows that system throughput increases monotonically with the target SNR, indicating that improved link quality can effectively support higher-order modulation schemes and higher data rates in the U6GHz band with 400 MHz bandwidth scenario. In the low-SNR region, the system mainly operates in the QPSK modulation region, where both downlink and uplink rates remain relatively low and reliable transmission is prioritized. As SNR increases in the 16QAM and 64QAM regions, the downlink throughput increases more significantly. In particular, after entering the 64QAM region at around 14 dB, the downlink throughput increases rapidly from approximately 0.5 Gbps to around 1 Gbps, demonstrating the spectral efficiency gain enabled by higher-order modulation. When the system further enters the 256QAM region, the downlink rate continues to increase, reaching nearly 1.5 Gbps, which indicates that the capacity advantage of the 400 MHz bandwidth can be fully exploited under high-SNR conditions. In contrast, the uplink throughput increases more gradually, with a maximum value of approximately 0.44 Gbps, which is significantly lower than that of the downlink. This reflects the asymmetry between downlink and uplink in terms of transmit power, antenna configuration, resource allocation, or link budget. Overall, the results indicate that the target SNR is a key factor affecting the U6GHz throughput performance. High-SNR conditions can substantially enhance the peak downlink capability, while uplink performance remains an important aspect for further optimization.

\section{CONCLUSION AND FUTURE DIRECTIONS}
\label{sec:VII}
\subsection{Conclusion}

This paper has presented a systematic review of spectrum allocation and standardization activities for sub-6 GHz and new mid-band frequencies, together with an overview of the spectrum planning strategies adopted by major countries and regions for future 6G systems. The wideband large-scale MIMO channel sounder designed for new mid-band measurements was also introduced. The propagation characteristics of four representative XL-MIMO architectures, namely co-located, cell-free, sparse, and movable XL-MIMO, were comprehensively investigated. The particular attention was given to their spatial non-stationarity, near-field propagation behavior, achievable capacity, and corresponding channel modeling approaches. In addition, recent advances in channel estimation, beamforming design, and artificial-intelligence-assisted signal processing were reviewed, providing a broad overview of the enabling techniques required for practical XL-MIMO deployment.

To further evaluate the potential of XL-MIMO in the new mid-band, systems equipped with 1536 and 768 antenna elements were comparatively analyzed. Field measurements conducted in the U6GHz band were used to assess the practical performance of the system and to examine several key performance indicators in realistic deployment scenarios. The results demonstrate that the target signal-to-noise ratio is a critical factor affecting system performance in the U6GHz band. A sufficiently high signal-to-noise ratio can significantly improve the peak downlink capacity and fully exploit the spatial multiplexing gain provided by large-scale antenna arrays. By contrast, uplink performance remains more constrained and requires further optimization in terms of link budget, channel estimation accuracy, power control, and transmission strategy. Overall, the findings confirm that new mid-band XL-MIMO is a promising technical solution for achieving enhanced coverage, high spectral efficiency, and large system capacity in future 6G networks.

 \subsection{Future Directions}
 
Nevertheless, further research is still required on measurement-based channel modeling, near-field and spatially non-stationary signal processing, low-complexity transceiver design, uplink enhancement, and practical array deployment. These issues will be essential for translating the theoretical advantages of XL-MIMO into robust and commercially deployable 6G systems.

Continuously strengthen the field verification of 6-24 GHz typical frequency band. Conduct multi-frequency field tests in dense urban areas, low-altitude scenarios, and typical vertical industry scenarios, and form datasets, model parameters, and evaluation methods that can be used for standardization and industrial verification. Combine with key technologies such as XL-MIMO, organize system-level experiments, focusing on verifying coverage continuity, edge rate, beam robustness, energy efficiency, site reuse capability, and construction cost, to provide quantitative basis for frequency planning.

Make early preparations for the coexistence of satellite-ground spectrum and the verification of mobile phone direct connection to satellites. Focus on Sub-3 GHz, S/L frequency bands, U6GHz, new mid-band, and Ku/Ka frequency bands, and conduct research on satellite-ground link budget, co-frequency/adjacent-frequency interference, dynamic spectrum sharing, beam isolation, power control, geographical isolation, and terminal access mechanism.


\bibliographystyle{IEEEtran}
\bibliography{ref}


\end{CJK}
\end{document}